\documentclass[groupedaddress,superscriptaddress,amsmath,amssymb]{revtex4}
\usepackage{float}
\usepackage{bbm}
\usepackage{bm}
\usepackage{graphicx,graphics}
\usepackage{soul}
\usepackage[makeroom]{cancel}
\usepackage{braket}
\usepackage{mathtools}
\usepackage{multirow}

\usepackage{algcompatible}
\usepackage{algorithm}
\usepackage{algpseudocode}
\usepackage{graphicx}

\usepackage[normalem]{ulem} 
\usepackage{cleveref}
\usepackage[svgnames]{xcolor}

\usepackage{amsmath}
\usepackage{amsthm}

\newcommand{\rS}{\mathrm{S}}
\newcommand{\rE}{\mathrm{E}}

\begin{document}

\title{Controlling quantum many-body systems using reduced-order modelling}

\author{I. A. Luchnikov}
\email{luchnikovilya@gmail.com}
\affiliation{Russian Quantum Center, Skolkovo, Moscow 143025, Russia}
\affiliation{National University of Science and Technology ``MISIS'', 119049 Moscow, Russia}

\author{M. A. Gavreev}
\affiliation{Russian Quantum Center, Skolkovo, Moscow 143025, Russia}
\affiliation{National University of Science and Technology ``MISIS'', 119049 Moscow, Russia}

\author{A. K. Fedorov}
\email{akf@rqc.ru}
\affiliation{Russian Quantum Center, Skolkovo, Moscow 143025, Russia}
\affiliation{National University of Science and Technology ``MISIS'', 119049 Moscow, Russia}


\begin{abstract}
Quantum many-body control is among most challenging problems in quantum science, due to computational complexity of related underlying problems. We propose an efficient approach for solving a class of control problems for many-body quantum systems, where time-dependent controls are applied to a sufficiently small subsystem. 
The approach is based on a tensor-networks-based scheme to build a low-dimensional reduced-order model of the subsystem's non-Markovian dynamics. Simulating dynamics of such a reduced-order model, viewed as a ``digital twin" of the original subsystem, is significantly more efficient, which enables the use of gradient-based optimization toolbox in the control parameter space. 
We validate the proposed method by solving control problems for quantum spin chains. In particular, the approach automatically identifies sequences for exciting the quasiparticles and guiding their dynamics to recover and transmit information. Additionally, when disorder is induced and the system is in the many-body localized phase, we find generalized spin-echo sequences for dynamics inversion, which show improved performance compared to standard ones. 
Our approach by design takes advantage of non-Markovian dynamics of a subsystem to make control protocols more efficient, and, under certain conditions can store information in the rest of the many-body system and subsequently retrieve it at a desired moment of time. We expect that our results will find direct applications in the study of many-body systems, in probing non-trivial quasiparticle properties, as well as in development control tools for quantum computing devices.

\end{abstract}

\maketitle

\section{Introduction}

The remarkable experimental capabilities have led to the advent of quantum technologies and inspired intense efforts to develop optimal control methods for quantum systems across several fields (for a review, see Ref.~\cite{Calarco2015}). Finding optimal control sequences for quantum many-body systems is a particularly important, but challenging task. Generally, full simulation of a many-body system dynamics for a given choice of control parameters requires resources exponential in the number of degrees of freedom; in gradient-based optimization methods over a large parameter space, such simulation has to be repeated many times, which makes such methods prohibitively demanding.

To overcome this challenge, gradient-free optimization methods combined with tensor-networks-based dynamics simulation were applied to control problems including many-body ground state preparation~\cite{caneva2011chopped, doria2011optimal, caneva2014complexity, van2016optimal}. Recently, methods using reinforcement learning techniques have been proposed~\cite{metz2022self, yao2021reinforcement}; such approaches can also be viewed as gradient-free optimization since they do not use the gradient of the reward function. The gradient-free optimization based control methods, however, are generally expected to be less efficient than gradient based methods~\cite{nocedal1999numerical}. Recently, Ref.~\cite{jensen2021achieving} demonstrated the advantage of gradient-based methods for the problem of ground state preparation by crossing the superfluid to Mott-insulator phase transition in the Bose-Hubbard model. 

Here, we propose a different approach to a class of problems where a sufficiently small subsystem $\mathcal S$ of a many-body system is subject to time-dependent controls. Focusing on time evolution of degrees of freedom in $\mathcal S$, we represent the rest of the many-body system $\mathcal E$ by its lower-dimensional ``twin'', or a {\it reduced-order}~\cite{brunton2022data, antoulas2005approximation, kim2022fast, luchnikov2021probing} model. Such a reduction effectively keeps track of  the relevant degrees of freedom in $\mathcal E$, discarding the ones which have little or no influence on dynamics of $\mathcal S$. 
The reduced-order model may involve effective Hilbert space dimension that is orders of magnitude smaller than the one in the original problem. This allows us to use the powerful toolbox of gradient-based optimization methods, using automatic differentiation techniques~\cite{liao2019differentiable} to calculate the gradient of the loss function. 

Practically, to build a reduced-order model, we employ tensor-network techniques, that are widely used for dimensionality reduction in quantum many-body physics~
\cite{bridgeman2017hand,Orus2019-3} and applied mathematics~\cite{oseledets2010tt, oseledets2011tensor}. Unifying the developed reduced-order modeling scheme with gradient based optimization yields an efficient method for quantum many-body control. 

We use our approach to automatically design protocols for manipulating information propagation in strongly interacting systems. First, we consider an one-dimensional (1D) XYZ quantum Heisenberg chain with extra fields that break integrability. For simplicity, we choose a single spin as the subsystem where time-dependent controls are applied. Our algorithm is able to find sequences that restore quantum information locally, or transmit to another end of the chain. Physically, the identified pulse sequences of local operations inject and re-absorb long-lived quasiparticles in an optimized way. 

Further, we apply the optimization method to the many-body localized phase. We are able to find control protocols for local dynamics inversion that outperform existing spin-echo-type protocols for many-body localized systems ~\cite{serbyn2014interferometric}. Thus, our approach enables automated discovery of optimal generalized spin-echo sequences in interacting systems.

The method describe here can be readily applied in experiments with the current generation of noisy, intermediate-scale quantum (NISQ) devices~\cite{Preskill2018,Aspuru-Guzik2021, doherty2013nitrogen}. Various quantum computing platforms, including programmable Rydberg simulators, trapped ions, isolated spin impurities in solids, and superconducting circuits arrays realize 1D spin chains~\cite{bernien2017probing, zhang2017observation, mazurenko2017cold, bermudez2010localization, barends2015digital, las2015fermionic, pasienski2010disordered, d2013quantum} with the possibility to control qubits individually by means of optical or microwave pulses. Our approach shows that the non-Markovianity of many-body environment can be employed for generating excitations and information spreading across the system. We expect that a modification of our approach may also be used for many-body state preparation. 

Here we are focused on the the realization of the control method in the coherent phase of quantum many-body systems assuming specific techniques to avoid fast thermalization. 
In the thermalized phase, our approach is not applicable and, at the same time, there are no reasons to expect here a possibility to maintain controllable coherent dynamics required in applications.

Our work is organized as follows.
In Sec.~\ref{sec:building_a_reduced_order_model}, we describe our general approach to building a reduced-order model.
We illustrate this methods for a quantum spin chain, our primary quantum many-body model of interest, in Sec.~\ref{sec:reduced-order-modeling-of-a-quantum-environment}.
In Subsec.~\ref{subsec:reduced-order-modeling-based_optimal_control}, 
we discuss the way to designing control protocols on the basis of reduced-order models.
We illustrate several control protocols using the proposed method:
in Subsec.~\ref{subsec:dynamics_inversion_via_optimal_control}, we demonstrate the inversion of dynamics of the system in time; 
in Subsec.~\ref{subsec:controllable_information}, we discuss controllable information propagation across the system, in particular we show a possibility to realize information transmission from Alice to Bob locating at different ends of the chain.
Finally, we conclude and discuss potential next steps in developing the proposed method in Sec.~\ref{conclusion}.

\medskip

\medskip

\section{Building a reduced-order model}
\label{sec:building_a_reduced_order_model}
\begin{figure*}
    \centering
    \includegraphics[width=1.\textwidth]{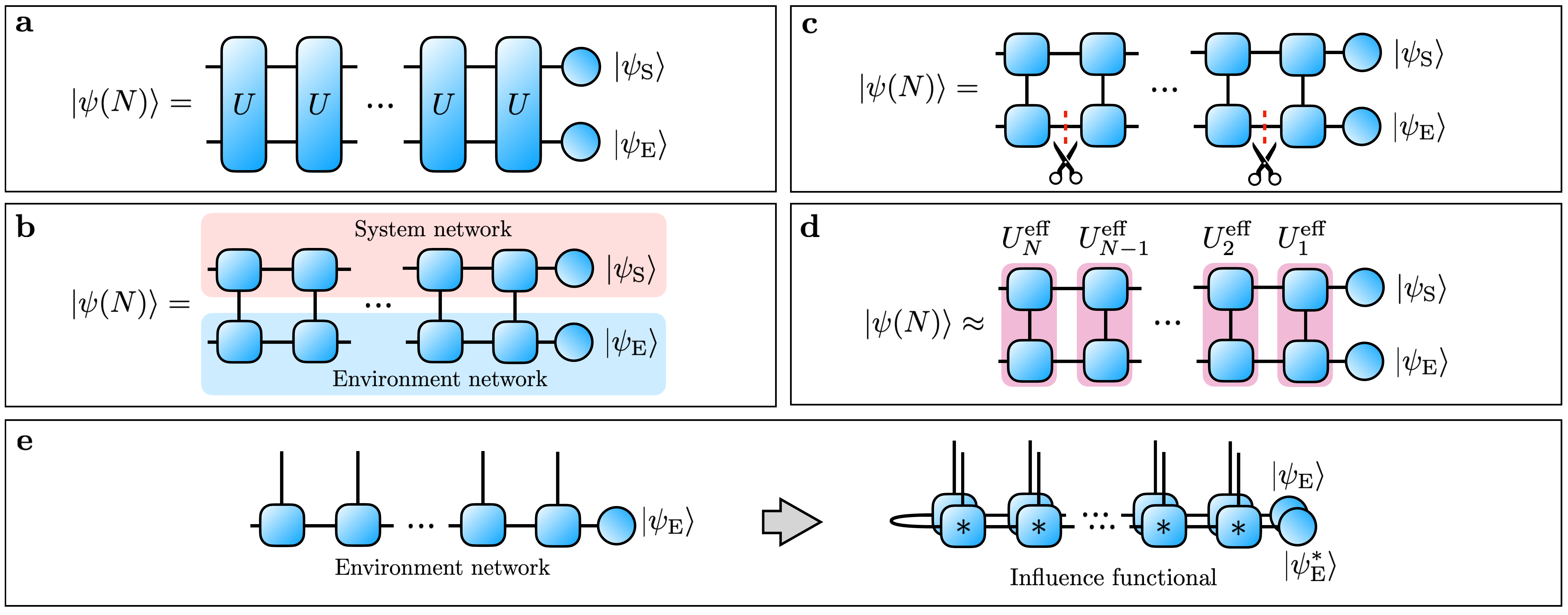}
    \caption{\textbf{a} Representation of a coupled system and environment dynamics as a tensor network. \textbf{b} The same tensor network after decomposition of each unitary block into two blocks. The entire network brakes up into two sub-networks namely a system network and an environment network shaded by red and blue regions correspondingly. \textbf{c} Schematic representation of the truncation procedure of the environment network. \textbf{d} The truncated tensor network that approximates the exact system and environment coupled dynamics. \textbf{e} Connection between the environment network and the influence functional. As one can see, the influence functional is a convolution of the environment network with complex conjugated version of itself.}
    \label{fig:environment_truncation_introduction}
\end{figure*}

Consider a many-body quantum system that consists of two parts. Assume, that the dynamics of the first part is of our interest, while the dynamics of the second part is not. This induces a natural separation of a many-body system into the target system (the first part) and the environment (the second part) in spirit of the theory of open quantum systems. Our first goal is to build a low-dimensional effective model of the target system dynamics, whose numerical simulation is much faster in comparison with the original model. This is the key step allowing one to run thousands of optimization iterations at the control signal adjustment stage. Since dynamics of the environment is not of out interest, one can reduce its dimension in such a way that dynamics of the target system remains almost the same. We utilize tensor network techniques for this purpose. We split the entire system dynamics into a sequence of unitary transformations $U$ and represent it as a tensor network shown in Fig~\ref{fig:environment_truncation_introduction}\textbf{a}.
Next, we use a singular-value decomposition (SVD) to decompose each $U$ that is seen as a tensor with four entries into two subtensors ending up with a decomposed tensor network shown in Fig~\ref{fig:environment_truncation_introduction}\textbf{b}. As one can note, the entire tensor network breakes up into two subnetworks, namely an {\it environment network} that is highlighted by the blue color and a {\it system network} that is highlighted by the red color. Note, that both networks have the form almost identical to a matrix product state (MPS) with only difference in the last dangling edge. The environment network by construction describes all the environmental effects in the system's dynamics. By reducing its dimension we end up with an effective low-dimensional environment network that leads to almost the same dynamics of the target system. We utilize the standard MPS truncation technique to build the effective low-dimensional environment network, this procedure schematically represented in Fig~\ref{fig:environment_truncation_introduction}\textbf{c}. We fix a desirable accuracy $\epsilon$ of the truncation and the truncation algorithm provides an effective environment network with bond dimension $r(k)$ depending on discrete time $k$. By taking the convolution between the effective environment network and the system network, as it is sketched in Fig~\ref{fig:environment_truncation_introduction}\textbf{d}, we get the effective low-dimensional model of the target system dynamics of time-depended dimension $d^{\rm eff}_k = d r(k)$, where $d$ is the target system dimension. The natural formulation of the truncation technique in terms of discrete in time open quantum dynamics is available in Appendix \ref{appx:open-quantum-dynamics-view}. Formal algorithm with its justification is given in Appendix \ref{appx:dimensionality_reduction_algorithm}.

It is worth to notice, that the environment network is closely connected to an {\it influence functional} \cite{feynman2000theory, breuer2002theory} that has been recently applied to a numerical simulation of quantum dynamics in variety of contexts. The most widely spread use case of the influence functional is the numerical simulation of a dynamics of an open quantum system coupled with a harmonic bath. Early approach to this problem \cite{makri1995tensor_theory, makri1995tensor_numerical} cuts off long-time memory effects by removing multipliers from the exact analytical form of the influence functional of a harmonic bath making it tractable for numerical treatment. More recent approaches such as TEMPO \cite{strathearn2018efficient} and its variations and improvements \cite{jorgensen2019exploiting, ye2021constructing} use low-rank tensor-network representation of the influence functional to improve accuracy and include long-time memory effects in the consideration. Another recent approach \cite{mascherpa2020optimized} aimed on replacement a complex harmonic bath by a simple one allowing numerical simulation of the system's dynamics. The core idea of the approach is to find such a  surrogate bath, which has the same two-time correlation functions. In case of a Gaussian bath this is equivalent to the equality of influence functionals. The combination of tensor network techniques with the theory of the influence functional were recently applied to develop new analytical and numerical methods for correlated spin systems dynamics analysis and simulation. In particular, self-consistency equation for the influence functional has been introduced in \cite{lerose2021influence, sonner2021influence, lerose2022overcoming} which allows one to study long-time thermalized dynamics of spin systems both analytically and numerically. It has been shown, that influence functional admits exact disorder averaging \cite{sonner2022characterizing} making it possible to study many-body localization (MBL) phases rigorously.

Approaches based on the influence functional are especially successful in case of irreversible processes with weak memory effects. In these cases temporal entanglement is weak and a low-rank MPS can efficiently approximate an influence functional. Whereas such systems are fundamentally very interesting, they are not well controllable due to the high information loss rate. Systems with long memory effects and weak information loss are better controllable and more interesting from the optimal control perspective. For describing environmental effects in such kind of problems the environment network suits better. To justify this claim let us consider the connection of the environment network with the influence functional that is represented in Fig.~\ref{fig:environment_truncation_introduction}\textbf{e}. As one can see, the environment network with bond dimension $r$ corresponds to the influence functional with bond dimension $r^2$. This means that the environment network can describe higher temporal entanglement and more complex memory effects. But on the other hand, the environment network can not efficiently describe irreversible loss of information. Any loss comes at the cost of increasing bond dimension. Therefore, environment network suits well for describing processes that are well controllable, i.e. with small information loss and high temporal entanglement and complements the influence functional based approaches. This motivates our choice of the environment network for the optimal control purposes. 

\section{Reduced-order modeling of a quantum spin chain}
\label{sec:reduced-order-modeling-of-a-quantum-environment}

\begin{figure*}[ht]
    \centering
    \includegraphics[scale=0.25]{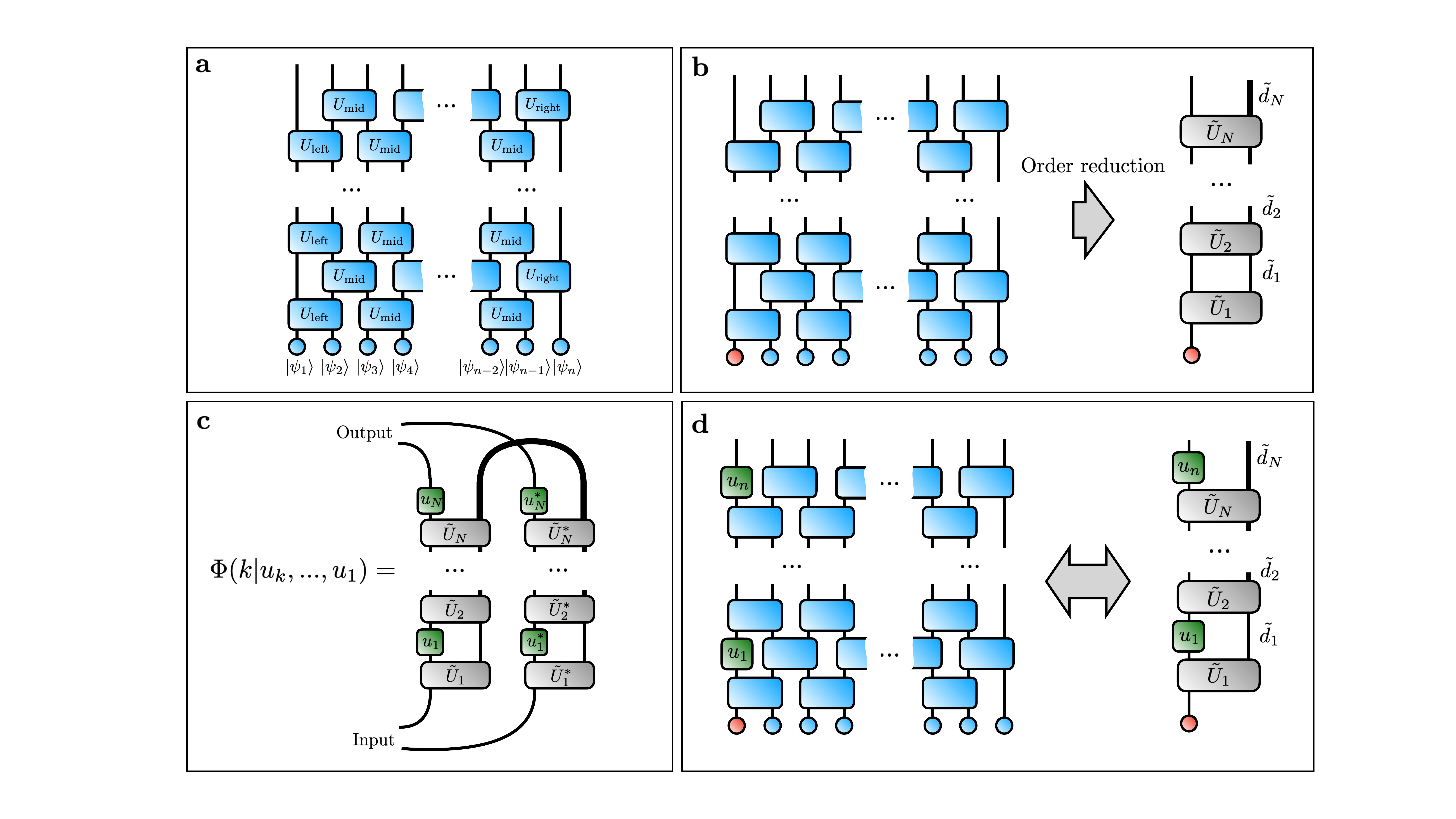}
    \caption{{\bf a} Tensor diagram representing the spin chain state after unitary evolution. Total number of spins is $n$, $N$ layers corresponds to the total evolution time $T = \tau N$. {\bf b} Tensor diagram illustrating the transition to the reduced-order model describing dynamics of the first spin. Varying edges thickness represents an increase of the effective environment dimension with time. The dimension of the joint system and effective environment state is depicted as $d^{\rm eff}_k$. {\bf c} Tensor diagram representing the quantum channel driving the dynamics of the target spin in terms of the reduced-order model. Asterisk symbol depicts complex conjugate. {\bf d} Tensor diagram showing how one applies control signal $\{u_1,\dots,u_N\}$ to the target spin (in the given case target spin is the first spin). The reduced-order model and the exact model can be used interchangeable, because they lead to the same response of the target spin to a control sequence.}
    \label{fig:main_tensor_diagrams}
\end{figure*}

We begin validation of the proposed reduced-order modeling technique from building a reduced-order model of a discretized in time XYZ quantum Heisenberg chain with external magnetic field and with open boundary conditions. The dynamics of a spin chain within this model is described by two-spin unitary operators that read
\begin{equation}
    U_{\#} = \exp\left(-i\tau H_{\#}\right),
\end{equation}
where $\# \in \{{\rm left}, {\rm mid}, {\rm right}\}$ and $\tau$ is a time step size. The corresponding two-spin Hamiltonians take the following form
\begin{eqnarray}
    H_{\rm left} = &&\sum_{k\in\{x, y, z\}}J_k \sigma^{k}_{0}\sigma^{k}_{1} + h_k \sigma^{k}_0 + \frac{h_k}{2} \sigma^{k}_1\nonumber \\
    H_{\rm mid} = &&\sum_{k\in\{x, y, z\}}J_k \sigma^{k}_{0}\sigma^{k}_{1} + \frac{h_k}{2} \sigma^{k}_0 + \frac{h_k}{2} \sigma^{k}_1\nonumber \\
    H_{\rm right} = &&\sum_{k\in\{x, y, z\}}J_k \sigma^{k}_{0}\sigma^{k}_{1} + \frac{h_k}{2} \sigma^{k}_0 + h_k\sigma^{k}_1,
\end{eqnarray}
where $\sigma^{k}_i$ denotes a $k$-th Pauli matrix ($k\in \{x, y, z\}$) acting on a spin number $i$, $J_k$ is a coupling constant, $h_k$ is a component of the external magnetic field.
In what follows we use tensor network diagrams in order to represent the final state of the spin chain after unitary evolution. One can think of $U_{\#}$ as a four-way tensor that is represented graphically as a block with four edges. Combining blocks $U_{\#}$ and initial spin chain's state into a tensor network, one represents the state of the spin chain at time $T=N\tau$ as it is shown in Fig.~\ref{fig:main_tensor_diagrams}{\bf a}, where $n$ is a number of spins, $N$ is a discrete time or a number of unitary layers, $\ket{\psi(0)} = \bigotimes_{i=1}^n\ket{\psi_i}$ is an initial state of the spin chain.
In the limit $\tau \rightarrow 0$, $\tau N \rightarrow T$ one restores the standard continuous in time XYZ quantum Heisenberg model in an external magnetic field.

As a target system we choose either the first or the middle spin of the spin chain and the rest of the spin chain we treat as the environment. Throughout the paper we use $l$ to denote the number of the target spin. For large spin chains the reduced-order modeling technique presented in Sec.~\ref{sec:building_a_reduced_order_model} is not directly applicable, since it requires explicit manipulations with the exact environment, whose dimension grows exponentially with number of subsystems. However, one can take advantage of the environment structure, i.e. in the given case it is either a single chain connected to the first spin or two chains connected to the middle spin. If environment has a chain-like structure, one can build the effective environment by iterative adding more subsystems to it and truncating it when necessary. Such an iterative approach does not require explicit manipulations with the complete environment and therefore scalable and applicable to large chain-like environments.

We perform such an iterative model order reduction procedure and end up with an effective low-dimensional model describing dynamics of the spin of interest. The transition to the effective model describing dynamics of the first spin is represented in terms of tensor diagrams in Fig.~\ref{fig:main_tensor_diagrams}{\bf b}, where by the red color we highlight the first spin whose dynamics is of our interest. Similar illustration can be build for the case of the middle spin. The dynamics of the target spin withing the reduced-order model reads
\begin{eqnarray}
    &&\ket{\psi^{\rm eff}(k)} = U^{\rm eff}_k\dots U^{\rm eff}_2U^{\rm eff}_1\ket{\psi_{l}},\nonumber \\
    &&\varrho_l(k) = {\rm Tr}_{\rm E}\left(\ket{\psi^{\rm eff}(k)}\bra{\psi^{\rm eff}(k)}\right),
\end{eqnarray}
where $l$ is the number of the target spin (first/middle), $\ket{\psi^{\rm eff}(k)}$ is the joint target spin and effective environment state at discrete time $k$ and $\varrho_l(k)$ is the state of the target spin. Note, that  in general case the dimension $d^{\rm eff}_k = 2r(k)$ of $\ket{\psi^{\rm eff}(k)}$ increases with time. An explanation of this effect is that the target spin gets entangled with more spins of environment with time, and therefore one needs to include more degrees of freedom of the environment into consideration.

\begin{figure*}[ht]
    \centering
    \includegraphics[scale=0.35]{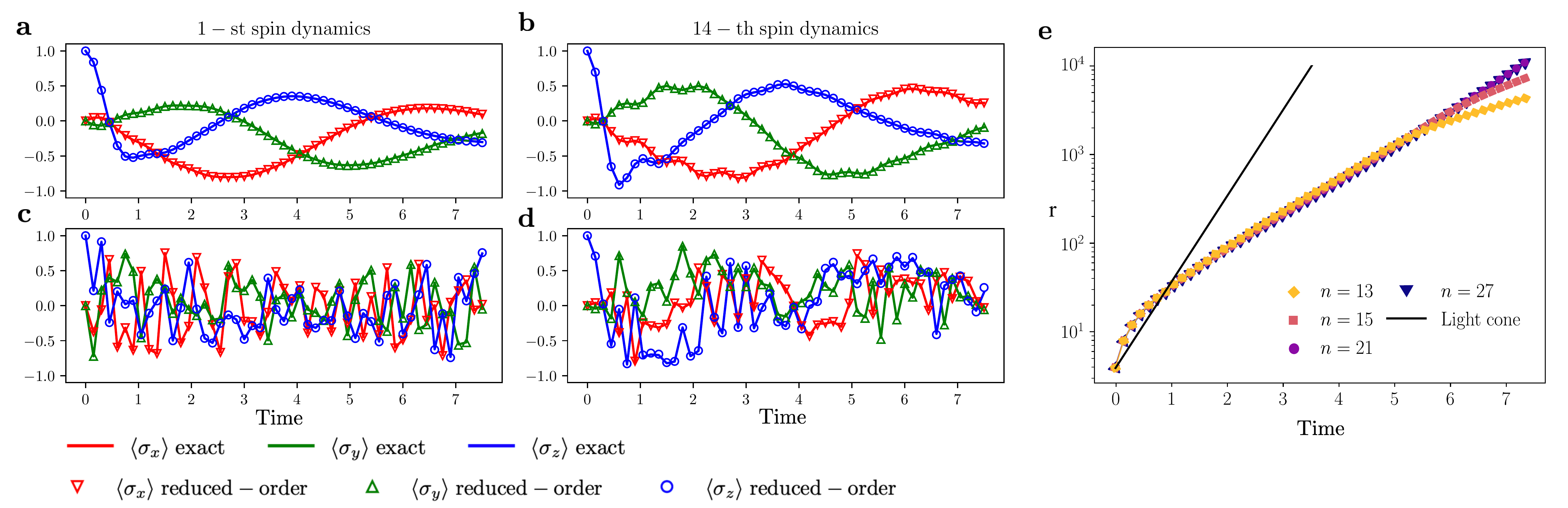}
    \caption{{\bf a} Comparison of the exact and the reduced-order model based dynamics of the first spin in the discrete in time XYZ model with external field consisting of $27$ spins. The total number of discrete time steps is $N=50$ that corresponds to total simulation time $T=7.5$. $\langle \sigma _i\rangle = {\rm Tr}(\sigma_i \varrho)$, where $i \in \{x, y, z\}$. {\bf b} All else being equal for the $14$-th (middle) spin as the target system. {\bf c} All else being equal for a random control signal. {\bf d} All else being equal for a random control signal and $14$-th (middle) spin as the target system.
    {\bf e} Comparison of the reduced-order model dimension with the light cone based estimation (black line) of the effective dimension for different numbers of spins $n$. $\epsilon = 0.01$ is an accuracy threshold used to build a reduced-order model (see Appendix~\ref{appx:dimensionality_reduction_algorithm} for the formal definition of $\epsilon$).
    }
    \label{fig:dynamics_prediction}
\end{figure*}

We compare the exact dynamics of the target spin with the dynamics simulated by use of the reduced-order model. We choose the following parameters of the model $J_x = 0.9$, $J_y = 1$, $J_z = 1.1$, $h_x = 0.2$, $h_y = 0.2$, $h_z = 0.2$, $\tau = 0.15$, and the following initial states of the target spin and the environment
\begin{equation}
    \ket{\psi_{\rm S}} = \ket{\uparrow},\quad \ket{\psi_{\rm E}} = \bigotimes_{i=1}^{n-1}\ket{\downarrow},
\end{equation}
where the number of spins is $n=27$, and the chosen model parameters correspond to a weakly non-integrable dynamics in the continuous in time case. We built the effective environment network by iterative adding spins to it,
and truncating it each time when its dimension exceeds $r_{\rm max} = 512$. We set the truncation accuracy (see Appendix \ref{appx:open-quantum-dynamics-view} and \ref{appx:dimensionality_reduction_algorithm} for more details) to be $\epsilon = 0.01$. The comparison of the exact target spin dynamics simulation with the dynamics simulation based on the reduced-order model is given in Fig.~\ref{fig:dynamics_prediction}{\bf a}, {\bf b}. 
One can note that the dynamics of the reduced-order model matches perfectly the exact one.

In order to demonstrate that the reduced-order model is capable of the external control signal response prediction, we apply a random control signal (a sequence of random one-qubit unitary gates) to the target system and compare its exact dynamics with the reduced-order model based one. The comparison is given in Fig.~\ref{fig:dynamics_prediction}{\bf c}, {\bf d}. As before, one can see that the exact and the reduced-order model based dynamics match each other.

In order to study how the dimension of the reduced-order model, i.e. $2r(k)$, scales with time for various numbers of spins and demonstrate the effect of dimensionality reduction we plot it in Fig.~~\ref{fig:dynamics_prediction}{\bf e} against time for different $n$ and the target spin fixed to be the first spin.


We also use the exact simulation of the whole spin chain dynamics in order to estimate the number of spins that are covered by the light cone propagating from the first spin. This number shows how many spins are involved in the dynamics of the first spin and thus the dimension of the Hilbert space of those spins is an upper bound of the reduced-order model's dimension. We plot how this upper bound evolves with time in Fig.~\ref{fig:dynamics_prediction}{\bf e}. One can see that the dimension of the reduced-order model grows much slower compared to the light cone based estimation of the effective dimension. While the reduced-order model's dimension for $27$ spins reaches $\approx 10^4$ at the final time step, the light cone covers the entire spin chains, which means that all $26$ spins of the environment are involved in the first spin dynamics. This is an evidence of the proposed reduced-order modeling technique efficiency because the naive light cone based estimation of the effective dimension results in $d^{\rm eff}_N\approx2^{27}\approx 1.3 * 10^8$ while the reduced-order modeling technique results in $d^{\rm eff}_N\approx 10^4$. However, the model reduction soonly becomes intractable with increasing simulation time, since $r(k)$ grows exponentially, and one can not simulate thermalization of the target spin properly. Nevertheless, our goal is to build the reduced-order model suitable for further simplification of different control problems and for this purpose the proposed technique suits well.

In the following section we move forward and apply the developed reduced-order modeling scheme to various optimal control problems.

\section{Many-body optimal control: methodology and numerical experiments}
\label{sec:methodology-and-numerical-experiments}

\subsection{Reduced-order modeling based optimal control}
\label{subsec:reduced-order-modeling-based_optimal_control}
The developed reduced-order modeling technique gives rise to a new class of optimal control methods in quantum many-body physics. Indeed, the main difficulty towards efficient quantum many-body optimal control is the necessity of running dynamics simulation thousands of times. This difficulty is substantially mitigated via the reduced-order modeling. The overall optimal control scheme breaks up into two steps:
\begin{enumerate}
    \item One builds a reduced-order model of a quantum many-body system. Now dynamics simulation can be run thousands of times within a reasonable time;
    \item One formulates an optimal control problem as an optimization problem written in terms of the reduced-order model and resolves this optimization problem using some optimization method.
\end{enumerate}
It remains unclear what kind of optimization method to use. A typical control problem written in terms of the optimization problem takes the following form:
\begin{eqnarray}
\label{eq:typical_optimization_problem}
    &&\underset{\{u_i\}_{i=1}^{\Delta N}}{\rm minimize} \ {\cal L}(u_1,\dots, u_{\Delta N}),\nonumber \\
    &&{\rm subject \ to} \ u_i^\dagger u_i = \mathbbm{1},
\end{eqnarray}
where ${\cal L}$ is the loss function that is written in terms of the reduced-order model and it measures how good a control signal is (the smaller value of ${\cal L}$ is, the better control signal is), $\{u_i\}_{i=1}^{\Delta N}$ is a $\Delta N$-steps sequence of unitary control gates applied to the system, i.e. it is a control signal that needs to be optimized, $\mathbbm{1}$ is the identity operator. Note, that Eq.~\eqref{eq:typical_optimization_problem} is the constrained optimization problem. Since control gates are unitary, an optimization technique of our choice must preserve $u_i^\dagger u_i = \mathbbm{1}$ for all gates. 
To solve the given optimization problem we found a Riemannian optimization algorithm \cite{boumal2020introduction, absil2009optimization} namely Riemannian ADAM optimizer \cite{becigneul2018riemannian, li2020efficient} to be efficient. It performs a gradient-based search of the optimal point on a manifold defined by the constraints, in our case on the manifold of unitary matrices (a special case of the complex Stiefel manifold \cite{edelman1998geometry, luchnikov2021riemannian, luchnikov2021qgopt, hauru2021riemannian}). We calculate gradient of $\cal L$ w.r.t $\{u_i\}_{i=1}^{\Delta N}$ utilizing the automatic differentiation technique \cite{liao2019differentiable}. Riemannian ADAM optimizer performs descent procedure towards the optimal point on the manifold of unitary matrices until convergence evaluating the gradient of ${\cal L}$ typically ten thousands times.  Note, that without the use of the reduced-order model, even a single calculation of the gradient becomes extremely memory demanding since automatic differentiation requires to keep all intermediate data in memory.

Let us consider a simple example. Suppose we are allowed to control the first spin of a spin chain. The transition to the reduced-order model in this case is schematically represented in Fig.~\ref{fig:main_tensor_diagrams}{\bf d}. The exact model and the reduced-order one are interchangeable in terms of the control response prediction. Suppose the optimal control goal is to have the initial and the final state of the first spin the same for any initial state. In this case the loss function can be written as ${\cal L}(u_1,\dots, u_N) = \|\Phi(N|u_1,\dots, u_N) - {\rm Id}\|^2_F$, where ${\rm Id}$ is the identity channel, $\Phi(N|u_1,\dots, u_N)$ is the channel that maps the initial first spin state to the final state. Here $\Phi(N|u_1,\dots, u_N)$ can be written in terms of the reduced-order model as it is shown in a tensor diagram Fig.~\ref{fig:main_tensor_diagrams}{\bf c} and thus the loss function is cheap to evaluate.

Below we provide three concrete examples of protocols for quantum control based on our approach.

\subsection{Dynamics inversion via optimal control}
\label{subsec:dynamics_inversion_via_optimal_control}

Here we consider the first quantum many-body optimal control problem. Suppose that one has access to a disordered spin system in the many-body localized (MBL) phase~\cite{abanin2019colloquium, ponte2015many, schreiber2015observation, slagle2016disordered} and it is allowed to apply a control signal to a dedicated single spin. One needs to design such a control protocol that runs dynamics of this spin ``backward'' in time. A typical example of such a protocol is a spin echo protocol \cite{serbyn2014interferometric, serbyn2013local} that runs dynamics ``backward'' in time in a sense that the controlled spin recovers information lost in the rest of the spin system after the spin flip operation. Our goal is to design an alternative control protocol that leads to better information recovery.

We start from a brief introduction into the origins of the spin echo protocol in MBL systems. Following the works \cite{serbyn2013local, huse2014phenomenology} this effect can be explained by use of the phenomenological model of the MBL phase. The MBL phase in the thermodynamic limit can be characterized by an infinite number of local integrals
of motion, which can be thought of as effective spin-half operators $\tau_i^z$. In this terms the MBL Hamiltonian takes the following form
\begin{eqnarray}
    \label{eq:phenom_mbl_hamiltonian}
    H_{\rm MBL} &&= \sum_i \tilde{h}_i \tau_i^z + \sum_{ij} {\cal J}_{ij} \tau^z_i\tau^z_j + \sum_{ijk} {\cal J}_{ijk} \tau^z_i\tau^z_j\tau^z_k + \dots,
\end{eqnarray}
where couplings ${\cal J}_{ij},\ {\cal J}_{ijk},\dots$ fall off exponentially with separation
with a characteristic localization length $\xi$. All terms of this Hamiltonian commute with each other. Therefor the total evolution operator $U_{\rm MBL}(t) = \exp\left(-i tH_{\rm MBL}\right)$ factorizes into the product of commuting exponents of individual terms. Consider one of those exponents that includes the participation of the first spin, it takes the following form:
\begin{eqnarray}
    &&\exp\left(-it{\cal J}_{1, i_2, \dots, i_m} \tau_{1}^z\tau_{i_2}^z\dots\tau_{i_m}^z\right) = \cos\left(t{\cal J}_{1, i_2, \dots, i_m}\right) - i \sin\left(t{\cal J}_{1, i_2, \dots, i_m}\right)\tau_{1}^z\tau_{i_2}^z\dots\tau_{i_m}^z.
\end{eqnarray}
Taking into account the following relation:
\begin{eqnarray}
    &&\exp\left(-it{\cal J}_{1, i_2, \dots, i_m} \tau_{1}^z\tau_{i_2}^z\dots\tau_{i_m}^z\right) \tau_1^x =\tau_1^x \exp\left(it{\cal J}_{1, i_2, \dots, i_m} \tau_{1}^z\tau_{i_2}^z\dots\tau_{i_m}^z\right).
\end{eqnarray}
which follows from the Pauli algebra, one finally ends up with
\begin{equation}
    U_{\rm MBL}(t) \tau_1^x U_{\rm MBL}(t) \tau_1^x = U_{{\rm MBL}/1}(2t),
\end{equation}
where $U_{{\rm MBL}/1}$ denotes the MBL evolution operator that describes the MBL dynamics of all spins but first spin and acts trivially (as the identity operator) on the first spin. It means that if the MBL system evolves for some time $t$ then one applies the spin-flip control gate $\tau_x$ to the first spin, then the system evolves for the same time $t$ again, and finally one applies the spin-flip control gate $\tau_x$ to the first spin again, one ends up with the completely the same state of the first spin as its initial state, i.e. the recovery of the information about the initial state of the first spin takes place. This is the essence of the spin echo protocol. The same consideration is valid for an arbitrary spin from the system.

However, the phenomenological model Eq.~\eqref{eq:phenom_mbl_hamiltonian} works well for the deeply localized phase. For the weakly localized phase, spin echo may barely be observed. Nevertheless, one can use the reduced-order modeling based optimal control technique from Subsection~\ref{subsec:reduced-order-modeling-based_optimal_control} to design an alternative multistep spin echo protocol suitable for a weakly localized phase. The multistep spin echo protocol consists in application of a sequence of unitary gates $\{u_{1},\dots,u_{\Delta N}\}$ instead of a single $\sigma_x$ gate, where $\Delta N$ is the duration of the protocol (number of control gates), to the target spin at the middle of the dynamics observation. We designed this protocol for one of the models experiencing MBL dynamics. The dynamics of this model is driven by the following Floquet operator \cite{sonner2021thouless}
\begin{equation}
    F = \exp\left[i\sum_{i=1}^{n-1} h_i \sigma^z_i + J \sigma_i^z\sigma_{i+1}^z\right]\exp\left[iJ\sum_{i=1}^n\sigma^x\right],
\end{equation}
where per-spin magnetic fields are random and sampled from the uniform distribution ${\rm Uniform}(0, \ 2\pi)$. The state of the whole system at discrete time $k$ reads $\ket{\psi(k)} = F^k \ket{\psi(0)}$. It is known that this model is in the localized phase for $J < J^*\approx 0.4$ \cite{sonner2021thouless}. In our numerical experiments we consider system consisting of $n=21$ spins, with coupling $J = 0.3$ corresponding to the localized phase, the target spin being the middle/first spin and compare the spin echo based dynamics inversion with a multistep spin echo based dynamics inversion designed by the proposed technique. We set a particular quenched disorder, i.e we picked a particular configuration of external magnetic fields from the distribution ${\rm Uniform}(0, \ 2\pi)$. We slightly generalized the spin echo protocol in order to make it better suitable for a particular quenched disorder. Instead of the instant swap of the spin by $\sigma_x$ at the middle of observation we apply an instant unitary gate $u$ that is optimized to achieve the best performance by using the proposed method. In other words, the generalized version of the spin echo protocol is the multistep spin echo protocol of duration $\Delta N = 1$. As the initial state of the environment (all spins but the target spin) we take $\ket{\psi_{\rm E}(0)} = \bigotimes_{i=1}^{n-1}\ket{\downarrow}$. For the total number of discrete time steps $N = 151$ we built a reduced-order model describing the dynamics of the target spin. For the multistep spin echo protocol we turn control ``ON'' in the time interval from $k_{\rm start} = 50$ to $k_{\rm stop} = 101$, i.e. the total protocol duration is $\Delta N = 51$ discrete time steps. For the spin echo protocol we turn control ON only for the single discrete time moment $k = 76$. 

To adjust the control signal for getting the best echo effect at the end of the dynamics, one needs to formulate the control problem as the optimization problem. The initial and the final states of the target spin are connected via the quantum channel $\Phi(N|u_1, \dots, u_{\Delta N})$ that can be defined by means of the reduced-order model. The closer $\Phi(N|u_1, \dots, u_{\Delta N})$ to some unitary channel is, the better echo effect one has. The mutual information $I(N|u_1, \dots, u_{\Delta N})$ between subsystem of the corresponding Choi matrix $\Omega(N|u_1, \dots, u_{\Delta N})$ reaches its maximum when $\Phi(N|u_1, \dots, u_{\Delta N})$ is a unitary channel (see Appendix \ref{appx:information_flow_visualization} for more details). Thus, maximizing $I(N|u_1, \dots, u_{\Delta N})$ one maximizes the echo effect. Therefore, the solution of the following optimization problem provides the optimal control signal
\begin{eqnarray}
    &&\underset{\{u_{1},\dots,u_{\Delta N}\}}{\rm maximize \ } I(N|u_{1},\dots,u_{\Delta N}),\nonumber\\
    &&{\rm subject \ to \ } u_i^\dagger u_i = \mathbbm{1}.
\end{eqnarray}
This optimization problem is solved by using the technique from Subsection~\ref{subsec:reduced-order-modeling-based_optimal_control}.

\begin{figure*}[ht]
    \centering
    \includegraphics[scale=0.5]{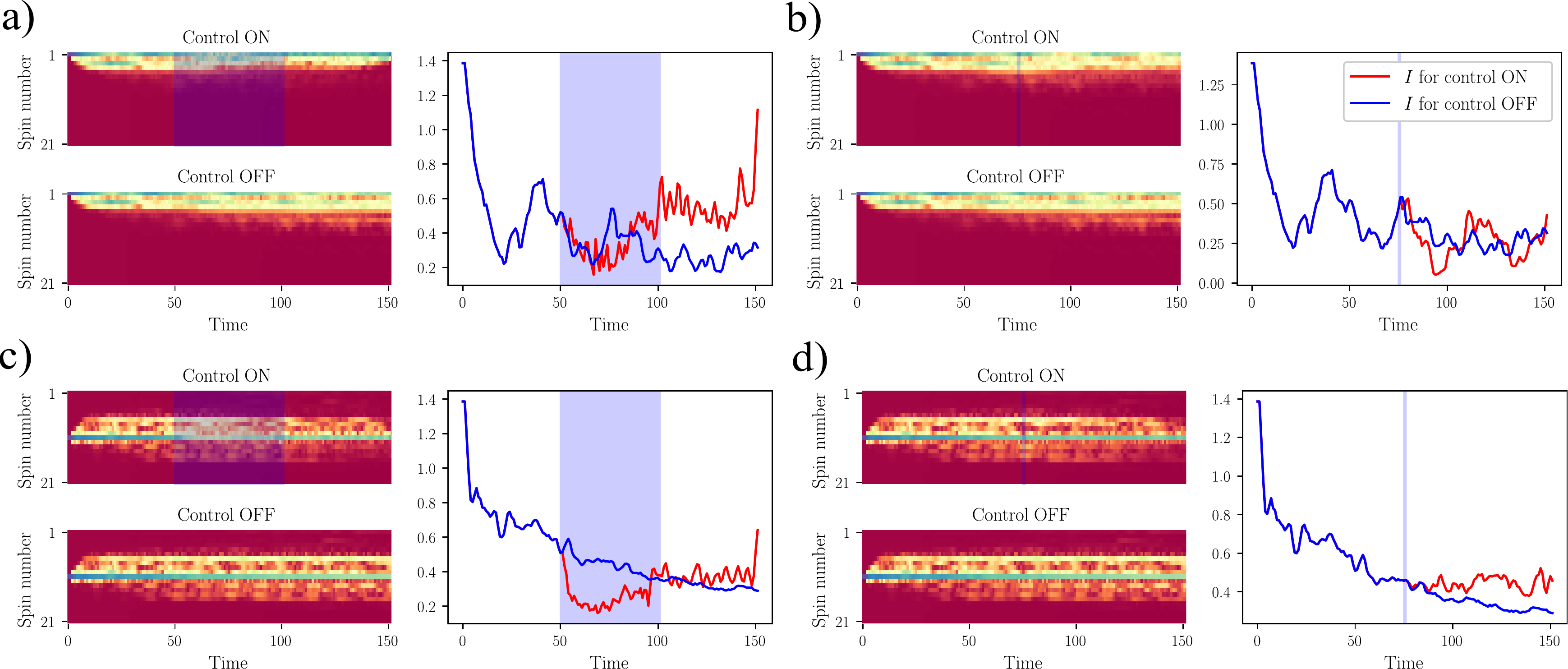}
    \caption{Information dynamics visualization for $J=0.3$ and {\bf a} first spin being the target spin and the multistep spin echo protocol, {\bf b} first spin being the target spin and the spin echo protocol, {\bf c} middle spin being the target spin and the multistep spin echo protocol, {\bf d} middle spin being the target spin and the spin echo protocol. Density plots represent how the rescaled mutual information $\log(I_{l\rightarrow m}(k) + 10^{-2})$ propagates across the spin chain, i.e. they show the flow of information about the initial state of the target spin. The blue band represents a region in time domain when the optimal control is applied. Line plots show how $I_{l\rightarrow l}$ evolves in time for the turned ON/OFF control protocol. The rise of $I_{l\rightarrow l}$ corresponds to the revival of information at the end of the evolution.}
    \label{fig:mbl_information_dynamics}
\end{figure*}

After getting the optimal control sequence, for both protocols we also run exact simulation of the entire spin chain in order to study the information flow under control and compare protocols with each other and with the case of control absence. Using the results of the exact simulation, we visualized information flow showing how the information about the initial state of the target spin spreads across the spin chain. For this purpose we utilize mutual information $I_{l\rightarrow m}(k)$ introduced in Appendix~\ref{appx:information_flow_visualization} that shows how much information about the initial state of $l$-th (target) spin is kept in $m$-th spin at discrete time moment $k$. We also separately plotted $I_{l\rightarrow l}(k)$ in order to demonstrate information revivals of the target spin. The results are given in Fig.~\ref{fig:mbl_information_dynamics}.
The three main conclusions could be made out of the Fig.~\ref{fig:mbl_information_dynamics}. First of all, we observe the information revival at the end of the evolution for both spin echo and multi step spin echo protocols. This means, that information about the initial target spin state is being reconstructed at the end of the evolution. Second, by looking on the density plots we note, that at the second half of the evolution information about the initial target spin state starts to propagate backward towards the target spin for both control protocols. This means that the dynamics inversion takes place. Finally, one can see that the multistep spin echo protocol outperforms the spin echo protocol in terms of revival amplitude. Therefore, the multistep spin echo protocol works better than the standard spin echo protocol.

\begin{figure*}[ht]
    \centering
    \includegraphics[scale=0.5]{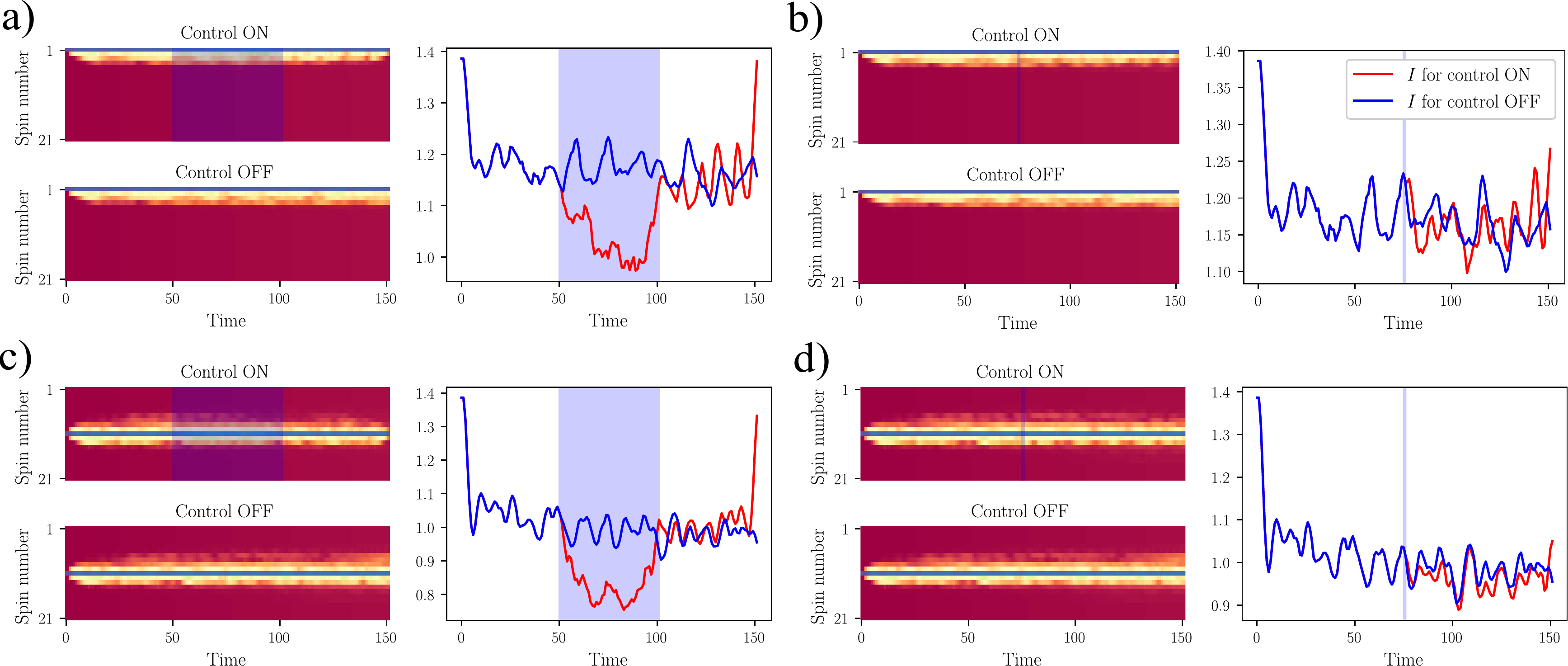}
    \caption{Averaged information dynamics visualization for $J=0.2$ and {\bf a} first spin being the target spin and the multistep spin echo protocol, {\bf b} first spin being the target spin and the spin echo protocol, {\bf c} middle spin being the target spin and the multistep spin echo protocol, {\bf d} middle spin being the target spin and the spin echo protocol. Density plots represent how the rescaled averaged mutual information $\log(\langle I_{l\rightarrow m}(k)\rangle + 10^{-2})$ propagates across the spin chain, i.e. they show the flow of information about the initial state of the target spin. The blue band represents a region in time domain when the optimal control is applied. Line plots show how $\langle I_{l\rightarrow l} \rangle$ evolves in time for the turned ON/OFF control protocol. The rise of $\langle I_{l\rightarrow l}\rangle$ corresponds to the revival of information at the end of the evolution.}
    \label{fig:averaged_mbl_information_dynamics}
\end{figure*}

To check that the conclusions above are still valid after averaging over disorder, we performed averaging over ten different disorder realizations for $J=0.2$ and all else parameters being the same. The same plots but for averaged quantities are show in Fig.~\ref{fig:averaged_mbl_information_dynamics}.
One sees that all the features we observed for a particular disorder realization are also valid in average.

\subsection{Controllable information propagation in a quantum spin chain}
\label{subsec:controllable_information}

\begin{figure*}[ht]
    \centering
    \includegraphics[scale=0.55]{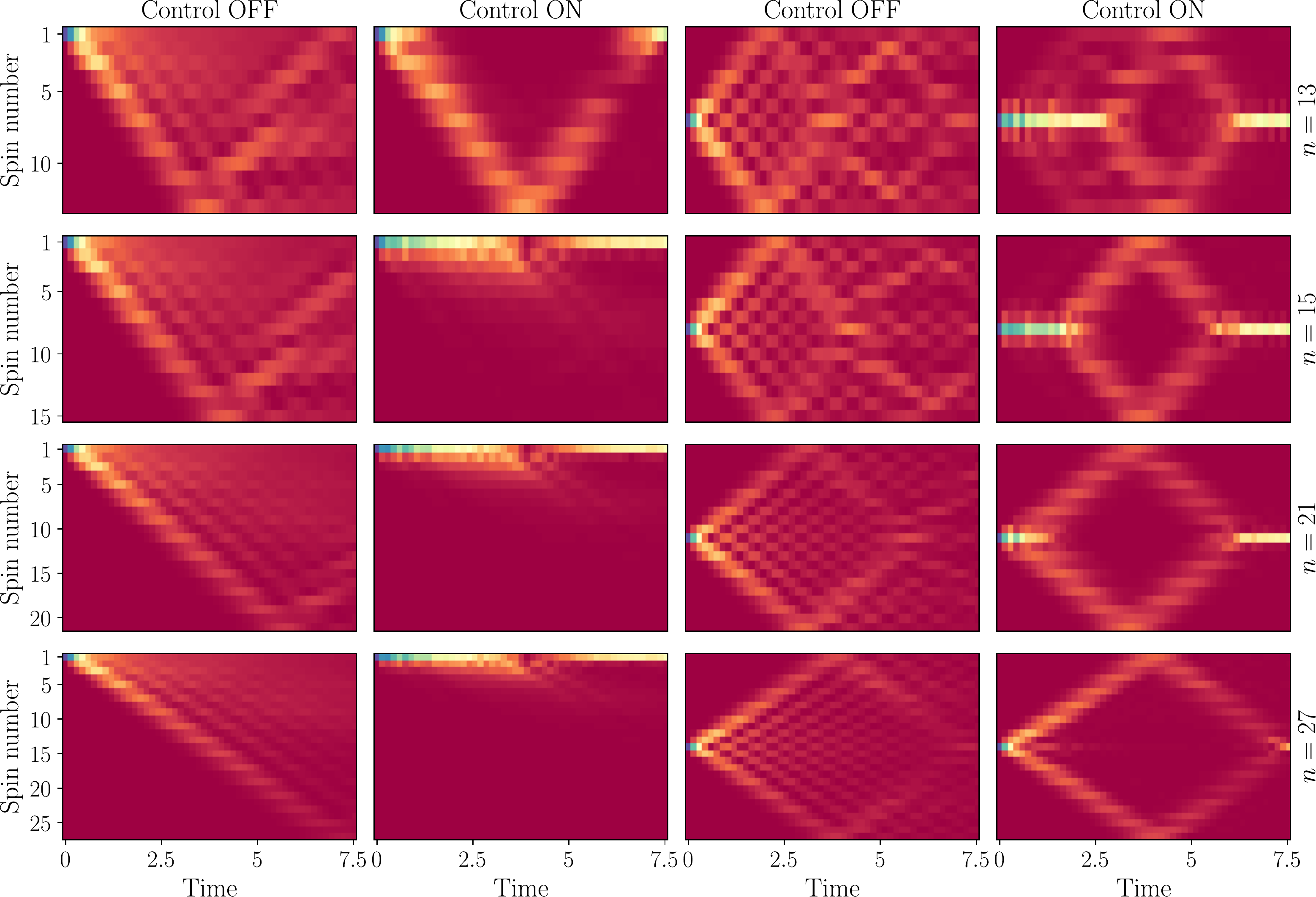}
    \caption{Density plots representing dynamics of $I_{l\rightarrow m}(k)$ with turned OFF/ON control ($x$-axis is time, $y$-axis is the number of a spin) for different values of $n$ and $l$ being equal to $1$ (the first spin) and $\frac{n-1}{2}+1$ (the middle spin).}
    \label{fig:XYZ_under_control}
\end{figure*}

In this subsection, we apply the proposed control technique to the control of information propagation in the discretized in time XYZ model discussed in Sec.~\ref{sec:reduced-order-modeling-of-a-quantum-environment}. We pick all the same parameters of the model as in Sec.~\ref{sec:reduced-order-modeling-of-a-quantum-environment} with the number of spins ranging from $13$ to $27$ and consider two control tasks aimed on controllable propagation of information through the spin chain. Within the first task we chose a bit artificial but complicated control problem causing non-trivial information flow under optimal control. The problem is formulated as follows: one needs to find such a control sequence $\{u_1,\dots,u_N\}$ applied to the target spin that $\varrho_l(0) = \varrho_l(N)$ and $\varrho_l(N/2) = \frac{I}{2}$, where $\varrho_l$ is the density matrix of the target spin. In other words, we want the information about the initial state of the target spin to be completely absorbed by the environment at time $T / 2$ and completely reconstructed back at the end of the dynamics. This control problem has the following formulation in terms of optimization:
\begin{eqnarray}
\label{eq:optimization_problem_1}
    &&\underset{\{u_i\}_{i=1}^N}{\rm minimize} \ \left\|\Phi\left(N/2|u_{N/2},\dots,u_{1}\right) - \Delta \right\|_F^2 + \left\|\Phi\left(N|u_{N},\dots,u_{1}\right) - {\rm Id}\right\|_F^2,\nonumber\\
    && {\rm subject \ to} \  u_i^\dagger u_i = \mathbbm{1}, \ \text{for all} \ i\in\{1,\dots, N\},
\end{eqnarray}
where $\Phi\left(k|u_k,\dots,u_{1}\right)$ is a quantum channel that maps the initial state of a target spin to the state of the first spin at discrete time $k$, ${\rm Id}$ is the identity quantum channel, $\Delta$ is a quantum channel that maps any state to the completely mixed state $\frac{1}{2}I$. We resolved this optimization problem using the technique from Subsection~\ref{subsec:reduced-order-modeling-based_optimal_control}. As before, we also did the exact dynamics simulation under the optimal control and without control in order to study how the information about the initial state of the target spin propagates in the spin chain. The information flow in all cases is visualized in Fig.~\ref{fig:XYZ_under_control}.
One can see that the optimal control sequence achieves the desired information flow, i.e., at the intermediate time, information about the initial state of the target spin dissolves in the rest of the spin chain; however, at the end of the dynamics, it is concentrated back in the target spin.

Interestingly, the optimal control sequence uses reflection of the information flow from borders of the spin chain as a resource when it is possible, i.e. when the information flow has enough time to reflect from a border and get back. Note, that this is a non-trivial effect of many body echo, that is recognized and utilized by the optimization algorithm with only use of the reduced-order model.

\begin{figure*}[ht]
    \centering
    \includegraphics[scale=0.43]{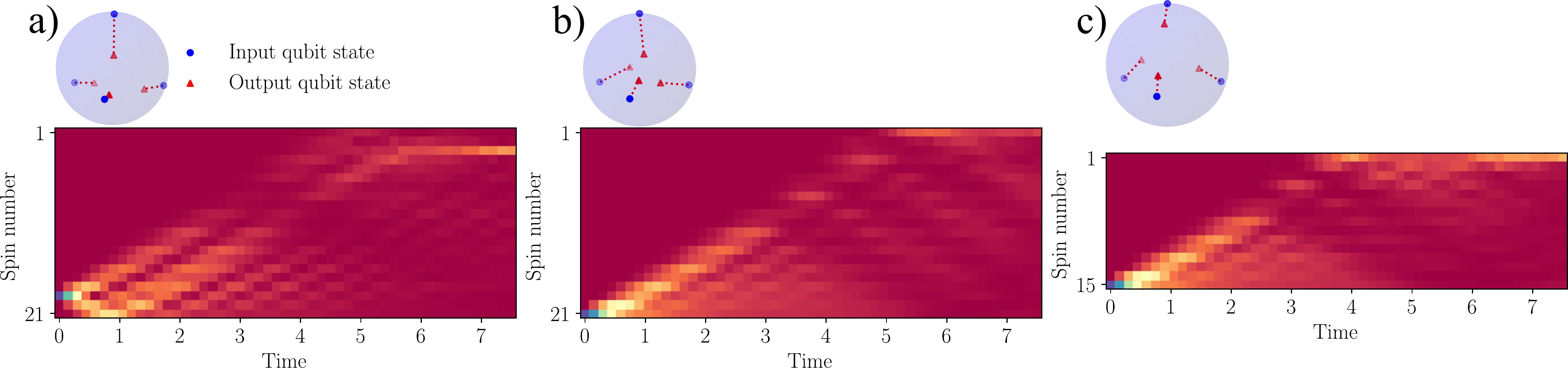}
    \caption{Density plots represent how information about Bob's spin propagates through the spin chain. Points in the Bloch ball show input states (Bob's spin initial states) and output states (Alice's spin final state). Plots correspond to the following parameters of the problem: {\bf a}) $n=21$, Bob's spin being the third spin Alice's spin being the spin number $19$; {\bf b}) $n=21$, Bob's spin being the first spin Alice's spin being the last {\bf c}) $n=15$, Bob's spin being the first spin Alice's spin being the last spin; spin;}
    \label{fig:XYZ_quantum_state_passing}
\end{figure*}

Within the second task, we apply the proposed method to design a control protocol allowing one to transfer quantum information through a spin chain. Let us assume that Bob prepares one of the spins in some state. The goal of Alice, who has access to one of another spins, is to apply such a control sequence $\{u_N,\dots,u_1\}$ to her spin, that after time $T$ Alice has her spin in the state as close as it is possible to the initial state of the Bob's spin. In other words, Alice has to ``catch'' information propagating from the Bob's spin and reconstruct the state of Bob's spin from this information. To formulate this task as an optimization problem, let us fix four linearly independent initial quantum states of the Bob's spin $\{\ket{\phi_i}\bra{\phi_i}\}_{i=0}^3$ whose corresponding Bloch vectors lie at the vertices of the tetrahedron, i.e. $\ket{\phi_i}\bra{\phi_i} = \frac{1}{2}\left(I + \sum_k s^{(i)}_k \sigma_k\right)$, where $s^{(i)}_k$ are components of vectors $\left\{\mathbf{s}^{(i)}\right\}_{i=0}^3$ that read
\begin{eqnarray}
    &&\mathbf{s}^{(0)} = (0, 0, 1), \ \mathbf{s}^{(1)} = \left(\frac{2\sqrt{2}}{3}, 0, -\frac{1}{3}\right), \nonumber \\ && \mathbf{s}^{(2)} = \left(-\frac{\sqrt{2}}{3}, \sqrt{\frac{2}{3}}, -\frac{1}{3}\right), \nonumber \\ &&\mathbf{s}^{(3)} = \left(-\frac{\sqrt{2}}{3}, -\sqrt{\frac{2}{3}}, -\frac{1}{3}\right).
\end{eqnarray}
Being able to pass these four states through the spin chain from Bob to Alice is enough to pass an arbitrary single spin state. For the fixed initial state of the Alice's spin (in our case $\ket{\downarrow}$) one can formulate the problem of transferring states $\{\ket{\phi_i}\bra{\phi_i}\}_{i=0}^3$ through the spin chain as the following optimization problem
\begin{eqnarray}
\label{eq:optimization_problem_2}
    &&\underset{\{u_i\}_{i=1}^N}{\rm minimize} \ \sum_{i=0}^3\|\ket{\phi_i}\bra{\phi_i} - \varrho\left(u_{N},\dots,u_1, \ket{\phi_i}\right)\|_F^2,\nonumber\\
    &&{\rm subject \ to} \ u_i^\dagger u_i = \mathbbm{1} \ {\rm for \ all} \ i\in{1, \dots, N},
\end{eqnarray}
where $\varrho\left(u_{N},\dots,u_1, \ket{\phi_i}\right)$ is the final state of the Alice's spin given the initial state of the Bob's spin and the control sequence. For each initial state $\ket{\phi_i}$ of the Bob's spin we build a separate reduced-order model describing dynamics of the Alice's spin and utilize it to compute $\varrho\left(u_{N},\dots,u_1, \ket{\phi_i}\right)$. The optimization problem Eq.~\eqref{eq:optimization_problem_2} as previous ones is solved by using the technique from Subsection~\ref{subsec:reduced-order-modeling-based_optimal_control}. In order to address the performance of the obtained optimal control sequence $\{u_N,\dots,u_1\}$ we compare initial states of the Bob's spin with final states of the Alice's spins and study how the information about the initial state of the Bob's spin propagates through the spin chain. The results are given in Fig.\ref{fig:XYZ_quantum_state_passing}.
One can see, that the optimal control sequence applied to the Alice's spin is able to partly reconstruct the initial state of the Bob's spin. One can also observe how Alice ``catches'' the light cone that propagates from the Bob's spin and preserves the information about Bob's spin up to the end of dynamics by using the optimal control sequence.

\section{Discussion and outlook}
\label{conclusion}

In the present paper, we have proposed a new method for many-body quantum control that is based on the reduced-order modeling scheme accelerating a numerical simulation of many-body quantum systems in many orders of magnitude. This acceleration makes it possible to run tens of thousands iterations of a gradient based control signal search in reasonable total time. We have validated the proposed method on number of control problems including controllable information spreading across a spin chain and dynamics inversion in the MBL phase. 

The proposed method gives rise to a new class of many-body control methods that have not been investigated before. Their field of applications varies from the development of new methods of error mitigation and noise suppression in quantum technologies to automatic search for new quantum materials, phases of matter and collective effects in many-body physics.

The proposed method can be generalized in various ways. For instance, instead of the iterative scheme for building the effective environment proposed in the paper, one can use tensor networks renormalization techniques such as introduced in Refs.~\cite{hauru2018renormalization, evenbly2015tensor, adachi2020anisotropic, harada2018entanglement, xie2012coarse}. They are not restricted by chain like environments and one can try to build reduced order models for 2D or even 3D many-body quantum systems and systems with irregular topology that are common in the field of quantum chemistry. Another possible generalization lies in the transition from the control of local observables and partial density matrices to macroscopic observables, e.g. total energy, total polarization, etc. Indeed, together with the environment dimensionality reduction one can ``renormalize'' macroscopic observables leading to reduced-order models of macroscopic observables dynamics. This opens new possibilities for steering quantum many-body systems between different phases of matter via external control. The transition from ``local'' to ``macroscopic'' is possible not only for observables but also for control signals. For instance, instead of applying a control signal to a single spin one may want to apply the same time-dependent magnetic field to all spins. In this case, design of the reduced-order model is definitely more involved, but with the great development of the tensor networks toolbox it may be possible. The next interesting generalization consists in extraction of a reduced-order model from observed experimental data. It is often the case that one has access to an experimental setup with possibility to measure the response of a quantum system of interest on an external control signal. The question is whether is it possible to build the reduced-order model of a system of interest in this case based purely on observed data? With use of algorithms such as tensor-train cross approximation~\cite{oseledets2010tt} one can try to do that efficiently and adaptively. Finally, the presented approach can be improved by unifying it with the influence matrix approach~\cite{lerose2021influence, lerose2022influence, sonner2021influence} allowing one to simulate long-time subsystems dynamics. The great development of tensor networks and dimensionality reduction techniques makes it possible to unify all the further generalizations of the proposed method into a universal framework opening great possibilities for automatic discovery of new quantum devices, phases of matter and quantum collective phenomena.

\section{Acknowledgments}

I.A.L., M.A.G., and A.K.F. acknowledge the support by the RSF Grant No. 19-71-10092 (studies of the many-body control approach) and the Priority 2030 program at the National University of Science and Technology ``MISIS” under the project K1-2022-027 (applications to spin chains).

\section{Code availability}
The code for all the numerical experiments is available via the link https://github.com/LuchnikovI/Quantum-many-body-dynamics-reduced-order-modeling. 

\appendix

\section{Building a reduced-order model: the theory of open quantum systems point of view}
\label{appx:open-quantum-dynamics-view}
Consider a system-environment Hilbert space, that reads
\begin{equation}
    {\cal H} = {\cal H}_\rS \otimes {\cal H}_\rE,
\end{equation}
where ${\cal H}_\rS$ is the $d_{\rm S}$-dimensional system Hilbert space and ${\cal H}_\rE$ is the $d_{\rm E}$-dimensional environment Hilbert space. For the sake of simplicity we will consider discretized time, however, the generalization of the proposed technique to the case of continuous time is also possible. The discrete in time dynamics of the joint system is driven by a unitary transformation $U:{\cal H}_\rS \otimes {\cal H}_\rE\rightarrow{\cal H}_\rS \otimes {\cal H}_\rE$, i.e. $\ket{\psi(N)} = U^N\ket{\psi(0)}$, where $\ket{\psi(0)}$ is an initial joint state of the system and environment. For the sake of simplicity $\ket{\psi(0)}$ is supposed to factorize as follows $\ket{\psi(0)} = \ket{\psi_\rS}\otimes \ket{\psi_\rE}$, nevertheless the generalization of the suggested approach to entangled initial states is straightforward. 

\begin{figure}[h]
    \centering
    \includegraphics[scale=0.17]{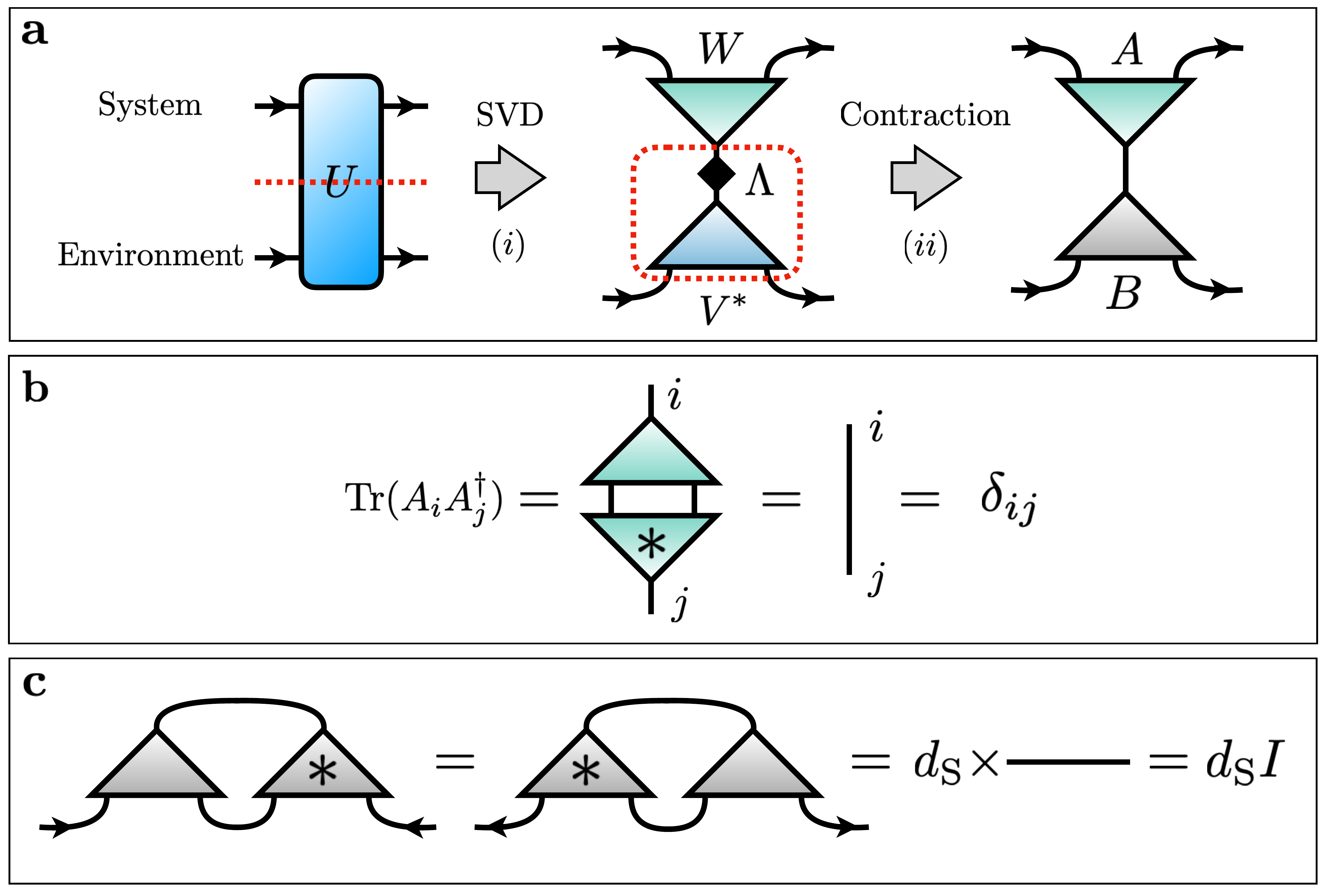}
    \caption{{\bf a} Two-step decomposition of $U$: ({\it i}) Singular value decomposition of $U$ w.r.t. the red dashed line separating $U$ into two parts; ({\it ii}) Contraction of the lower isometric block with the diagonal matrix $\Lambda$ with singular values into a single block. We additionally annotate tensor edges by arrows showing the input and the output of a unitary transformation. {\bf b} Orthogonality property for $A_i$. {\bf c} Orthogonality property for $B_i$.}
    \label{fig:u-decomposition}
\end{figure}

To separate system and environment for further environment dimensionality reduction, one needs to split $U$ into two parts. A unitary transformation $U$ can be represented as a $4$-way tensor. In terms of diagrammatic notations tensor $U$ is seen as a block with for edges. Let us introduce a dyadic decomposition $U = \sum_{i} A_i \otimes B_{i}$ that is defined via diagrammatic representation in Fig~\ref{fig:u-decomposition}{\bf a}.
Both objects $A_i$ and $B_i$ are seen as 3-way tensors, i.e. both have input and output ``physical'' indices and one index $i$ induced by the decomposition. Let us determine some useful properties of $A_i$ and $B_i$. First, one has the following orthogonality relation for $A_i$
\begin{equation}
\label{eq:orthogonality_of_A}
    {\rm Tr}(A_iA_j^\dagger) = \delta_{ij},
\end{equation}
where $\delta_{ij}$ is the Kronecker delta. Indeed, the core of the decomposition Fig.~\ref{fig:u-decomposition}{\bf a} is the singular value decomposition (SVD) and Eq.~\eqref{eq:orthogonality_of_A} follows directly from the definition of SVD. The diagrammatic representation of Eq.~\eqref{eq:orthogonality_of_A} is given in Fig.~\ref{fig:u-decomposition}{\bf b}. To determine orthogonality relations for $B_i$, we consider the following relation
\begin{equation}
    {\rm Tr}_{\rm S} (UU^\dagger) = {\rm Tr}_{\rm S} (I \otimes I) = d_{\rm S} I,
\end{equation}
where ${\rm Tr}_{\rm S}$ is the partial trace over the system, $I$ is the identity matrix. On the other hand, one can make use of the decomposition Fig.~\ref{fig:u-decomposition} and Eq.~\eqref{eq:orthogonality_of_A} and rewrite ${\rm Tr}_{\rm S} (UU^\dagger)$ as follows
\begin{equation}
    {\rm Tr}_{\rm S} (UU^\dagger) = \sum_{ij} {\rm Tr}_{\rm S} (A_iA_j^\dagger \otimes B_iB_j^\dagger) =\sum_{ij} {\rm Tr}(A_iA_j^\dagger) B_iB_j^\dagger = \sum_{ij} \delta_{ij} B_iB_j^\dagger = \sum_i B_iB_i^\dagger.
\end{equation}
Gathering all together one ends up with
\begin{equation}
\label{eq:B_orthogonality_1}
    \sum_i B_i B_i^\dagger = d_{\rm S} I.
\end{equation}
Considering ${\rm Tr}_{\rm S} (U^\dagger U)$ instead of ${\rm Tr}_{\rm S} (UU^\dagger)$, one also ends up with
\begin{equation}
\label{eq:B_orthogonality_2}
    \sum_i B_i^\dagger B_i = d_{\rm S} I.
\end{equation}
The diagrammatic interpretation of the expressions Eq.~\eqref{eq:B_orthogonality_1} and Eq.~\eqref{eq:B_orthogonality_2} is given in Fig.~\ref{fig:u-decomposition}{\bf c}. The most important consequence of these relations which we use further is that operators $K_i = \frac{1}{\sqrt{d_{\rm S}}}B_i$ form Kraus representation of a quantum channel, i.e. completely positive (CP) and trace-preserving (TP) map $\Psi$
\begin{equation}
\label{eq:environment_quantum_channel}
    \Psi[\cdot] = \sum_i K_i \cdot K_i^\dagger,
\end{equation}
where the CP property is guarantied automatically and $\sum_{i} K_i^\dagger K_i = I$ guaranties the TP property.

\begin{figure*}[ht]
    \centering
    \includegraphics[scale=0.25]{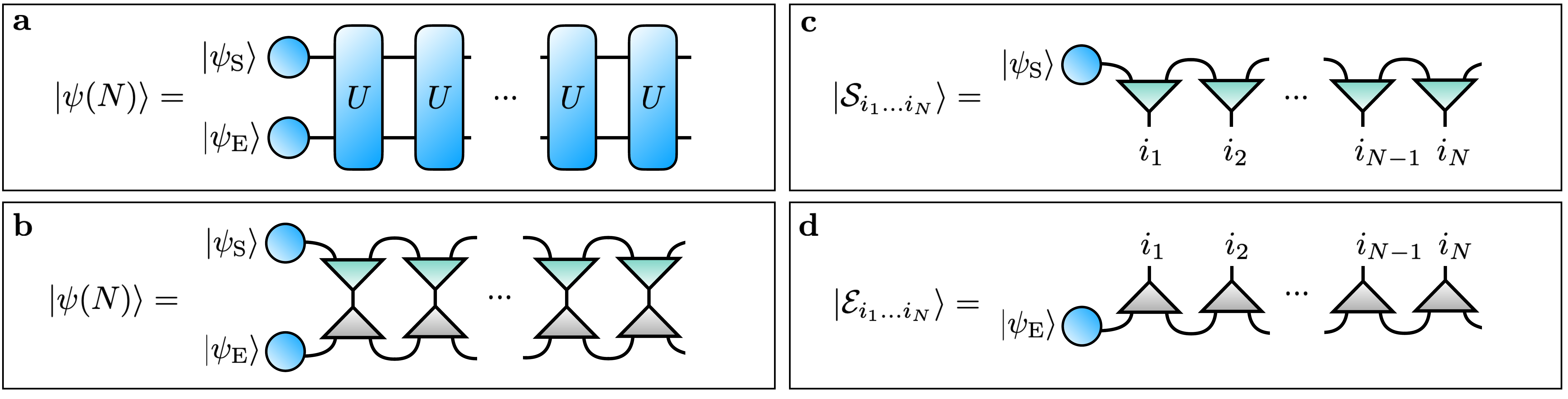}
    \caption{{\bf a} Tensor network representing the joint unitary dynamics of the system and the environment. {\bf b} The same tensor network with decomposed $U$ tensors according to Fig.~\ref{fig:u-decomposition}. {\bf c} System network. {\bf d} Environment network.}
    \label{fig:unitray_dynamics_decomposition}
\end{figure*}

Now let us use the decomposition Fig.~\ref{fig:u-decomposition}{\bf a} to represent the joint system and environment final state $\ket{\psi(N)} = U^N\ket{\psi(0)}$, where $N$ is the total number of discrete time steps, as a tensor network necessary to proceed with the environment dimensionality reduction. The straightforward representation in terns of a tensor network is given in Fig.~\ref{fig:unitray_dynamics_decomposition}{\bf a}. Applying the decomposition Fig.~\ref{fig:u-decomposition}{\bf a} to all $U$ tensors in Fig.~\ref{fig:unitray_dynamics_decomposition}{\bf a} one ends up with the tensor network Fig.~\ref{fig:unitray_dynamics_decomposition}{\bf b}. This tensor network can be splitted into two parts, a  system network $\ket{{\cal S}_{i_1\dots i_N}}$ and an environment network $\ket{{\cal E}_{i_1\dots i_N}}$, they read
\begin{eqnarray}
    &&\ket{{\cal S}_{i_1\dots i_N}} = A_{i_N}A_{i_{N-1}}\dots A_{i_1} \ket{\psi_{\rm S}}, \nonumber \\
    &&\ket{{\cal E}_{i_1\dots i_N}} = B_{i_N}B_{i_{N-1}}\dots B_{i_1} \ket{\psi_{\rm E}}.
\end{eqnarray}
The diagramatic representations of both networks are given in Fig.~\ref{fig:unitray_dynamics_decomposition}{\bf a}, {\bf b}.
The final joint system and environment state in terms of networks $\ket{{\cal S}_{i_1\dots i_N}}$ and $\ket{{\cal E}_{i_1\dots i_N}}$ can be written as follows
\begin{equation}
    \ket{\psi(N)}=\sum_{i_1,\dots,i_N} \ket{{\cal S}_{i_1\dots i_N}}\otimes \ket{{\cal E}_{i_1\dots i_N}}.
\end{equation}
Note, that both networks have a form very similar to the matrix product state (MPS) tensor network~\cite{orus2014practical, bridgeman2017hand}. The only difference with MPS is the additional dangling edge which is responsible for the final system (environment) state.

\begin{figure*}[ht]
    \centering
    \includegraphics[scale=0.28]{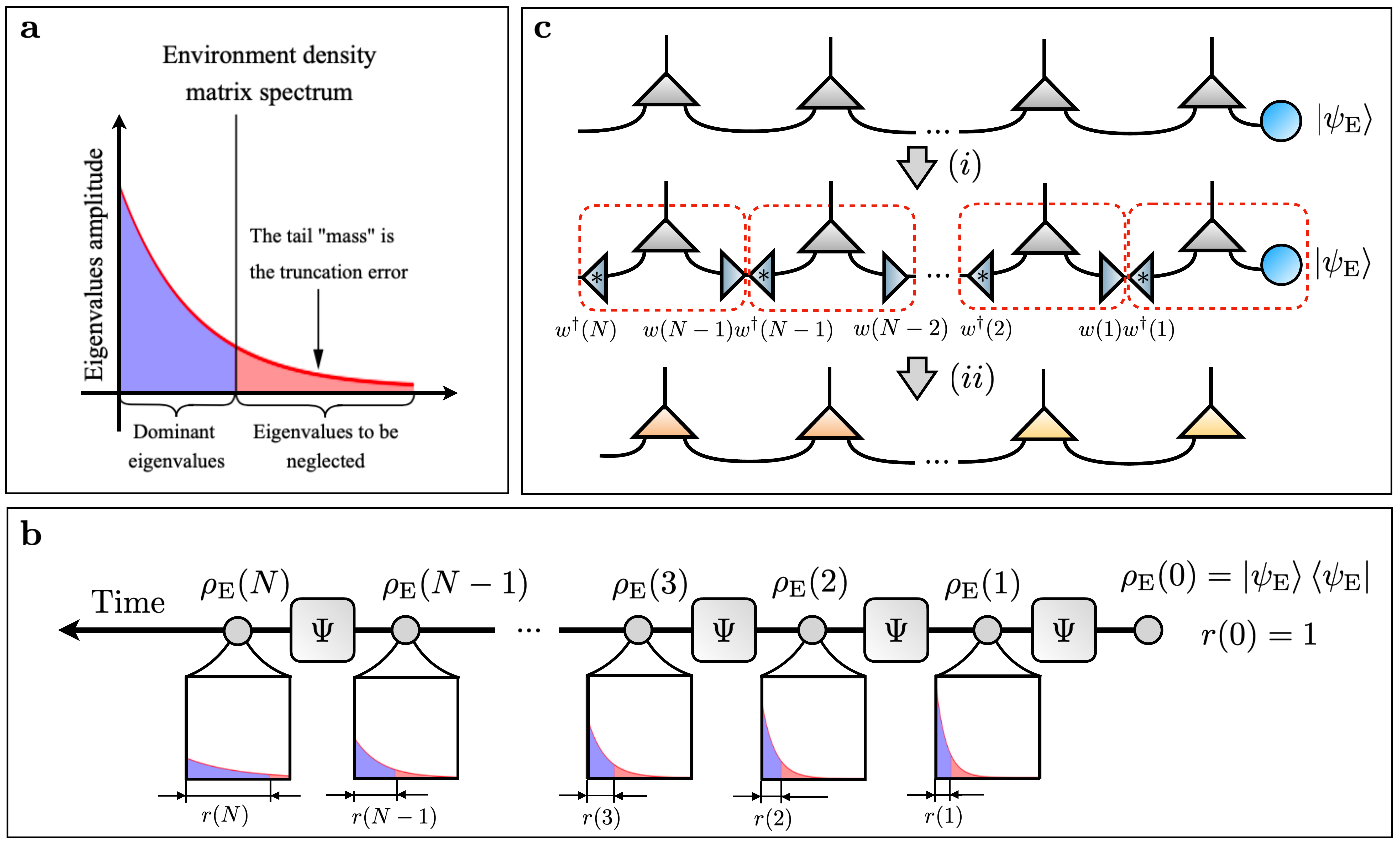}
    \caption{{\bf a} A typical spectrum of the environment density matrix. The tail of this spectrum contains only a tiny fraction of the total eigenvalues ``mass'' and thus can be truncated. {\bf b} A typical time evolution of the environment density matrix spectrum. It gets wider in time and therefore the principle subspace dimension $r(k)$ grows in time. {\bf c} The environment truncation scheme: ({\it i}) One inserts projectors $\omega(k)\omega^\dagger(k)$ on the principle subspace in between of blocks forming the environment network. ({\it ii}) One contracts tensors in dashed red boxes and ends up with new yellow blocks that form truncated environment network.}
    \label{fig:truncation_method}
\end{figure*}

At this point we are ready to perform environment dimensionality reduction. The object whose dimensionality is being reduced is the environment network. Due to the relation Eq.~\eqref{eq:B_orthogonality_1} the environment network is automatically in a so called left-canonical form, that is the starting point of the standard MPS truncation algorithm \cite{schollwock2011density, oseledets2011tensor, oseledets2010tt}. This gives rise to an efficient environment network truncation technique, that is equivalent to the standard MPS truncation algorithm. It is easy to formulate this technique purely in terms of the environment dynamics induced by the quantum channel $\Psi$ introduced in Eq.~\eqref{eq:environment_quantum_channel}. Consider discrete in time dynamics of the environment under the action of the quantum channel $\Psi$, i.e.
\begin{eqnarray}
    &&\varrho_{\rm E}(m) = \Psi[\varrho_{\rm E}(m-1)],\nonumber\\
    &&\varrho_{\rm E}(0) = \ket{\psi_{\rm E}}\bra{\psi_{\rm E}}.
\end{eqnarray}
The central object we care about is the spectrum $(\lambda_1(m), \lambda_2(m),\dots,\lambda_{d_{\rm E}}(m))$ of $\varrho_{\rm E}(m)$, where eigenvalues arranged in the non-ascending order. If the spectrum of $\varrho_{\rm E}(m)$ is mostly concentrated in $r(m)$ largest eigenvalues, then one can project $\varrho_{\rm E}(m)$ on a principle subspace that is the span of $r(m)$ dominant eigenvectors, i.e. eigenvectors with largest eigenvalues. This leads to the truncated version of the density matrix that reads
\begin{equation}
    \varrho_{\rm E}^{\rm trunc}(m) = \pi(m)\varrho_{\rm E}(m)\pi(m),
\end{equation}
where $\pi(m)$ is the orthogonal projector on the principal subspace.
To gain more intuition about how the principal subspace is determined, we schematically plotted a typical spectrum in Fig.~\ref{fig:truncation_method}{\bf a}. The relative error of the projection (truncation) reads
\begin{equation}
    \frac{\|\varrho_{\rm E}^{\rm trunc}(m) - \varrho_{\rm E}(m)\|_{F}}{\|\varrho_{\rm E}(m)\|_F} = \sqrt{\sum_{j=r(m)+1}^{d_{\rm E}} \lambda_j(m)},
\end{equation}
where $F$ stands for the Frobenius norm. In words it means that the error is equal to the square root of ``mass'' of eigenvalues in the spectrum tail that is cut and colored by red in Fig.~\ref{fig:truncation_method}{\bf a}. By establishing a desirable error threshold $\sqrt{\sum_{j=r(m)+1}^{d_{\rm E}} \lambda_j(m)}\leq \epsilon_m$ one determines the principal subspace dimension $r(m)$ and the principle subspace itself as the column space of the matrix $\omega(m)$ whose columns are $r(m)$ dominant eigenvectors. The principle subspace is seen as a low-dimensional effective environment Hilbert space at time $m$. Typically $\varrho_{\rm E}(m)$ gets nosier in time, i.e. its spectrum gets wider. In order to preserve the truncated spectrum tail ``mass'' the same, one needs to increase $r(m)$ with time. Therefore, $r(m)$ typically grows in time. This is schematically illustrated in Fig.~\ref{fig:truncation_method}{\bf b} where one can see how $r(m)$ grows with $m$ due to the widening of the spectrum. To obtain the truncated environment network it is enough to insert projection operators $\pi(m) = \omega(m)\omega^\dagger(m)$ in between of neighboring blocks $B_{i_{m+1}}$ and $B_{i_{m}}$ for all $m$ as it is shown in Fig.~\ref{fig:truncation_method}{\bf c}. This results in truncated blocks that read
\begin{equation}
    \widetilde{B}_i^{(m)} = \begin{cases}\omega(m) B_i \omega^\dagger(m-1), \ \text{if \ } m > 1, \\
    \omega(1) B_i \ket{\psi_{\rm E}}, \ \text{otherwise}, \end{cases}
\end{equation}
and in the truncated environment network $\ket{{\cal E}_{i_1\dots i_N}^{\rm trunc}} = \widetilde{B}_{i_N}^{(N)}\widetilde{B}_{i_{N-1}}^{(N-1)}\dots \widetilde{B}^{(1)}_{i_1}$. The error introduced by the whole procedure is bounded above as follows (see Appendix~\ref{appx:dimensionality_reduction_algorithm} and Ref.~\cite{oseledets2011tensor})
\begin{equation}
    \frac{\left\|\ket{{\cal E}^{\rm trunc}_{i_1\dots i_N}} - \ket{{\cal E}_{i_1\dots i_N}}\right\|_F}{\left\|\ket{{\cal E}_{i_1\dots i_N}}\right\|_F} \leq \sqrt{\sum_{m=1}^N \epsilon_m^2}.
\end{equation}
Therefore, if one require the error to be less or equal to some upper bound $\epsilon$ it is enough to set $\epsilon_m = \frac{\epsilon}{\sqrt{N}}$ which leads to
\begin{equation}
    \frac{\left\|\ket{{\cal E}^{\rm trunc}_{i_1\dots i_N}} - \ket{{\cal E}_{i_1\dots i_N}}\right\|_F}{\left\|\ket{{\cal E}_{i_1\dots i_N}}\right\|_F} \leq \epsilon.
\end{equation}
Varying the value of $\epsilon$ one achieves a trade off between accuracy of approximation and effective environment dimension $r(m)$.

Finally, having the truncated environment network, one can build the reduced-order model of the system dynamics. The reduced-order model is defined by ``effective'' gates $U_m^{\rm eff}$ of size $d_{\rm S}r(m) \times d_{\rm S}r(m-1)$ driving joint dynamics of the system and the effective environment, they read
\begin{equation}
    U_m^{\rm eff} = \sum_i A_i \otimes \widetilde{B}_i^{(m)}.
\end{equation}
Dynamics of the system with use of the reduced-order model can be calculated as follows
\begin{equation}
    \varrho_{\rm S} (k) = {\rm Tr}_{\rm E}\left(\prod_{m=1}^kU_m^{\rm eff}\ket{\psi_{\rm S}}\bra{\psi_{\rm S}}\left[\prod_{m=1}^kU_m^{\rm eff}\right]^\dagger\right).
\end{equation}
Note also, that it is allowed to apply arbitrary control gates to the system. The reduced-order model based prediction of the system dynamics under control would match the exact one.

\section{Building a reduced-order model: the algorithm and its justification}
\label{appx:dimensionality_reduction_algorithm}

In this appendix we justify the proposed environment network dimensionality reduction algorithm and provide its precise formulation. 

First, we discuss dimensionality reduction of the environment network at a specific discrete time moment. Let $\ket{{\cal E}^{(m)}_{i_1\dots i_N}}$ be the environment network whose dimension at a discrete time moment $m$ has been reduced, i.e.
\begin{eqnarray}
    &&\ket{{\cal E}^{(m)}_{i_1\dots i_N}}= B_{i_N}\dots B_{i_{m+1}}w(m)w^\dagger(m) B_{i_m} \dots B_{i_1} \ket{\psi_{\rm E}},
\end{eqnarray}
where $w(m)$ is a trial isometric matrix of size $d_{\rm E}\times r(m)$, $r(m)$ is a new environment dimension such that $r(m) < d_{\rm E}$. A natural choice of $w(m)$ is the one that leads to the minimal error, i.e. we require $w(m)$ to be the solution of the following optimization problem
\begin{eqnarray}
    &&\underset{w(m)}{\rm minimize}\left\|\ket{{\cal E}^{(m)}_{i_1\dots i_N}} - \ket{{\cal E}_{i_1\dots i_N}}\right\|_F^2,\nonumber\\
    &&{\rm subject \ to} \ w^\dagger(m) w(m) = I,
\end{eqnarray}
where the Frobenius norm is taken over all indices, i.e. ``physical'' index and the set of indices $\{i_1, \dots, i_N\}$.
The objective function ${\cal L}(\omega(m)) = \left\|\ket{{\cal E}^{(m)}_{i_1\dots i_N}} - \ket{{\cal E}_{i_1\dots i_N}}\right\|_F^2$ dramatically simplifies if one makes use of the property Eq.~\eqref{eq:B_orthogonality_2} and the introduced in Eq.~\eqref{eq:environment_quantum_channel} quantum channel $\Psi$. The simplified objective function takes the following form
\begin{equation}
    {\cal L}(\omega(m))= d_{\rm S}^N\left(1 - {\rm Tr}\left(w^\dagger(m) \varrho_{\rm E}(m) w(m)\right)\right),
\end{equation}
where $\varrho_{\rm E}(m) = \Psi^{m}[\ket{\psi_{\rm E}}\bra{\psi_{\rm E}}]$. Under the given constraints, this optimization problem is equivalent to the problem of finding $r(m)$ eigenvectors of $\varrho(k)$ corresponding to $r(m)$ maximal eigenvalues, i.e. the optimal $w(m)$ is the matrix whose columns are eigenvectors of $\varrho_{\rm E}(m)$ corresponding to $r(m)$ largest eigenvalues. In other words, the optimal orthogonal projector $\pi(m) = w(m)w^\dagger(m)$ is the projector on $r(m)$ leading eigenvectors of $\varrho(m)$.

Another question that arises here, is how can one determine $r(m)$? In practice, one usually has some desirable approximation accuracy (admissible error). Let us connect this accuracy and $r(m)$.
The relative error induced by the dimensionality reduction reads
\begin{eqnarray}
    \epsilon_m &&= \frac{\left\|\ket{{\cal E}^{(m)}_{i_1\dots i_N}} - \ket{{\cal E}_{i_1\dots i_N}}\right\|_F}{\left\|\ket{{\cal E}_{i_1\dots i_N}}\right\|_F} = \sqrt{\sum_{j={r(m)+1}}^{d_{\rm E}} \lambda_j(m)}.
\end{eqnarray}
The above relation for the error allows one to determine the minimal value of $r(m)$ that suits some desirable accuracy $\epsilon_m$ as a function $r(m) = f\left(\epsilon_m, \{\lambda_j(m)\}_{j=1}^{d_{\rm E}}\right)$ of eigenvalues and $\epsilon_m$. We do not provide a concrete form of $f$ here for the case of a single time step dimensionalite reduction but do this later for the case of all time steps dimensionality reduction. 

This scheme can be applied to the environment network multiple times leading to an algorithm allowing dimensionality reduction for all time steps. The overall algorithm reads
\begin{algorithm}[H]
\caption{Environment dimensionality reduction}
\begin{algorithmic}[1]
    \Require CPTP map $\Psi$, tensor $B_{i}$ forming the environment network, initial state $\ket{\psi_{\rm E}}$ of the environment, accuracy $\epsilon$ of the algorithm
    \Ensure The set of tensors $\left\{\widetilde{B}^{(m)}_i\right\}_{m=1}^N$ forming the reduced environment
    \State Set the initial density matrix of the environment $\varrho(0) = \ket{\psi_{\rm E}}\bra{\psi_{\rm E}}$ and the initial isometry $w(0)=\ket{\psi_{\rm E}}$
    \For{$m \gets 0$ to $n-1$}
        \State Propagate the density matrix of the environment forward in time: $\varrho(m+1) = \Psi[\varrho(m)]$
        \State Perform eigendecomposition of $\varrho(m+1)$: $U(m+1)\Lambda(m+1) U^\dagger(m+1) = \varrho(m+1)$
        \State Calculate the minimal environment dimension $r(m+1)$ that suits the given error upper bound $\epsilon$: $r(m+1) = f\left(\epsilon, \{\lambda_j(m+1)\}_{j=1}^{d_{\rm E}}\right)$
        \State Truncate the density matrix rank $\varrho(m+1)\gets U_{r(m+1)}(m+1) \Lambda_{r(m+1)}(m+1) U_{r(m+1)}^\dagger(m+1)$
        \State Calculate the next isometry $w(m+1)$: $w(m+1) = U_{r(m+1)}(m+1)$
        \State Calculate the next tensor $\widetilde{B}_i^{(m+1)}$: $\widetilde{B}_i^{(m+1)} = w^\dagger(m+1) B_i w (m)$
    \EndFor
\end{algorithmic}
\end{algorithm}
This algorithm results in the truncated environment network $\ket{{\cal E}_{i_1\dots i_N}^{\rm trunc}}$ that reads
\begin{equation}
    \ket{{\cal E}_{i_1\dots i_N}^{\rm trunc}} = w(N)\widetilde{B}_{i_N}^{(N)}\widetilde{B}_{i_{N-1}}^{(N-1)}\dots \widetilde{B}^{(1)}_{i_1} \ket{\psi_{\rm E}},
\end{equation}
Note, that one can omit $w(N)$ in the expression above. Indeed, we are not interested in the final state of the environment, we only care about the action of the environment on the system, i.e. the exact coincidence of environment networks is redundant, it is enough to have coincidence of  discretized Feynman-Vernon influence functionals \cite{feynman2000theory, sonner2021influence, lerose2021influence, lerose2022influence} that are easily expressed through environment networks
\begin{equation}
    \braket{{\cal E}_{j_1\dots j_N}^{\rm trunc}|{\cal E}_{i_1\dots i_N}^{\rm trunc}}\approx \braket{{\cal E}_{j_1\dots j_N}|{\cal E}_{i_1\dots i_N}}.
\end{equation}
Due to the property $w^\dagger(N)w(N) = I$ matrix $w(N)$ does not affect the value of $\braket{{\cal E}_{j_1\dots j_N}^{\rm trunc}|{\cal E}_{i_1\dots i_N}^{\rm trunc}}$
and can be safely omitted.

It is important to note, that the algorithm above is equivalent to the standard algorithm for MPS truncation \cite{schollwock2011density, oseledets2011tensor, oseledets2010tt}. The orthogonality relation $\sum_i B_i^\dagger B_i = d_{\rm S} I$ means that the environment network is in the left-canonical form that is the starting point of the standard MPS truncation algorithm. By forwarding the environment density matrix in discrete time via CPTP map $\Psi$ we push the orthogonality center from the right side to the left. The projection on the leading eigenvectors of the environment density matrix is equivalent to the SVD based truncation.

Finally, let us determine the function $f$. The error introduced by the entire algorithm is bounded above as follows \cite{oseledets2011tensor}
\begin{equation}
    \epsilon = \frac{\left\|\ket{{\cal E}^{\rm trunc}_{i_1\dots i_N}} - \ket{{\cal E}_{i_1\dots i_N}}\right\|_F}{\left\|\ket{{\cal E}_{i_1\dots i_N}}\right\|_F} \leq \sqrt{\sum_{m=1}^N \epsilon_m^2},
\end{equation}
where $\epsilon_m$ is the error of $m$-th time step dimensionality reduction.
Therefore, restricting the one-time-step dimensionality reduction error $\epsilon_m \leq \frac{\epsilon}{\sqrt{n}}$ we guarantee that the error introduced by the entire algorithm does not exceed $\epsilon$. This leads to the concrete form of the function $f$ guaranteeing a given accuracy $\epsilon$ of the algorithm
\begin{eqnarray}
    r(m) &&= f\left(\epsilon, \{\lambda_j(m)\}_{j=1}^{d_{\rm E}}\right) = d_{\rm E} - \sum_{r=1}^{d_{\rm E}}\eta\left(\frac{\epsilon}{\sqrt{n}}, \sqrt{\sum_{j=r+1}^{d_{\rm E}} \lambda_j(m)}\right),
\end{eqnarray}
where $\eta$ is defined as follows
\begin{equation}
    \eta(x, y) = \begin{cases}1,\quad \text{if} \ x > y, \\ 0, \quad \text{otherwise.}\end{cases}
\end{equation}

\section{Information flow visualization}
\label{appx:information_flow_visualization}

\begin{figure*}[ht]
    \centering
    \includegraphics[scale=0.27]{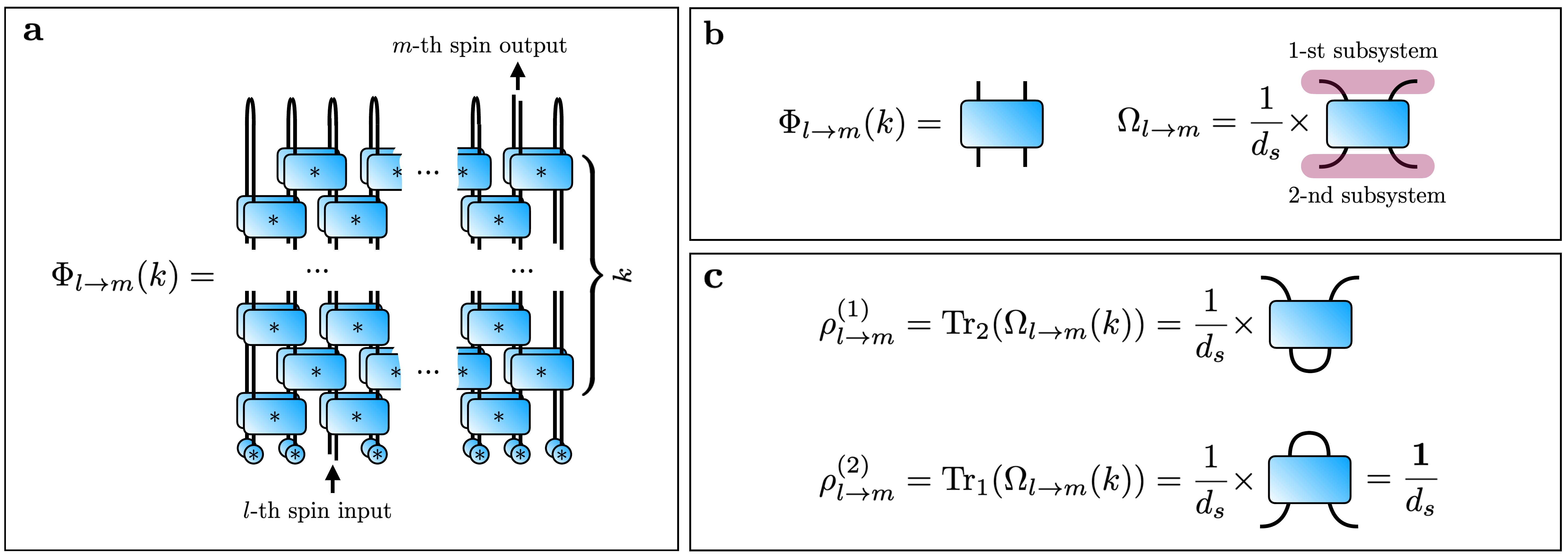}
    \caption{{\bf a} Diagrammatic representation of the quantum channel $\Phi_{l\rightarrow m}(k)$. {\bf b} Diagrammatic representation of the corresponding Choi matrix $\Omega_{l\rightarrow m}(k)$. Note, that the only essential difference between $\Phi_{l\rightarrow m}(k)$ and $\Omega_{l\rightarrow m}(k)$ is in the multiplier $\frac{1}{d_{\rm S}}$. {\bf c} Diagrammatic representation of $\varrho^{(1)}_{l\rightarrow m}(k)$ and $\varrho^{(2)}_{l\rightarrow m}(k)$. Note, that due to the TP property of $\Phi_{l\rightarrow m}(k)$, $\varrho^{(2)}_{l\rightarrow m}(k)$ is proportional to the identity operator.}
    \label{fig:mutual_information_explanation}
\end{figure*}

To validate optimal control results and gain intuition behind them, it is instructive to visualize how information about the initial state of a certain spin propagates across a spin chain. For this purpose, we introduce a quantum channel $\Phi_{l\rightarrow m}(k)$ that maps the initial state of $l$-th spin to the state of $m$-th spin at time $k$. Its diagrammatic representation is given in Fig.~\ref{fig:mutual_information_explanation}{\bf a}. This quantum channel fully characterizes correlations between the initial state of $l$-th spin and the state of $m$-th spin at $k$-th discrete time moment. To quantify correlations by a single value one can turn to the corresponding Choi matrix $\Omega_{l\rightarrow m}(k)$ that is represented in terms of tensor diagrams in Fig.~\ref{fig:mutual_information_explanation}{\bf b}. This Choi matrix is seen as the density matrix of two-component quantum system and thus the mutual information between those components $I_{l\rightarrow m}(k)$ is well defined and reads
\begin{eqnarray}
    I_{l\rightarrow m}(k) &&= S\left(\varrho^{(1)}_{l\rightarrow m}(k)\right) + S\left(\varrho^{(2)}_{l\rightarrow m}(k)\right) - S\left(\Omega_{l\rightarrow m}(k)\right),
\end{eqnarray}
where $S$ stands for Von Neumann entropy, $\varrho^{(1)}_{l\rightarrow m}(k)$ is the first component density matrix and $\varrho^{(2)}_{l\rightarrow m}(k)$ is the second component density matrix. Both $\varrho^{(1)}_{l\rightarrow m}(k)$ and $\varrho^{(2)}_{l\rightarrow m}(k)$ are represented in terms of tensor diagrams in Fig.~\ref{fig:mutual_information_explanation}{\bf c}. $I_{l\rightarrow m}(k)$ suits well for our visualization purposes, it shows how information about $l$-th spin propagates in discrete time $k$ and space $m$. Indeed, there are other quantities that may suit better for this role, e.g. quantum capacity~\cite{lloyd1997capacity, shor2002quantum, devetak2005private}, but we chose mutual information since it is easy to calculate.

\bibliography{bibliography.bib}
\end{document}